\documentclass[twocolumn,showpacs,preprintnumbers,amsmath,amssymb]{revtex4-1}

\usepackage{graphicx,latexsym}
\usepackage{amsmath,amssymb,amsfonts}
\usepackage{bm}
\usepackage{braket}
\usepackage{mathrsfs}
\newcommand{\up}{\uparrow}
\newcommand{\down}{\downarrow}
\newcommand{\m}{\mathcal}
\newcommand{\ep}{\varepsilon}

\usepackage[dvipsnames]{xcolor} 
\usepackage{color}
\begin{document}
\title{Emergence of Larkin-Ovchinnikov-type superconducting state in a voltage-driven superconductor}
\author{Taira Kawamura\email{tairakawa@keio.jp},$^{1,2}$ Yoji Ohashi,$^1$ and H.T.C. Stoof$^2$}
\affiliation{$^1$Department of Physics, Keio University, 3-14-1 Hiyoshi, Kohoku-ku, Yokohama 223-8522, Japan}
\affiliation{$^2$Institute for Theoretical Physics and Center for Extreme Matter and Emergent Phenomena, Utrecht University, Princentonplein 5, 3584 CC Utrecht, The Netherlands}
\date{\today}
\begin{abstract}
We theoretically investigate a voltage-biased normal metal-superconductor-normal metal (N-S-N) junction. Using the nonequilibrium Green's function technique, we derive a quantum kinetic equation, to determine the superconducting order parameter self-consistently. The derived equation is an integral-differential equation with memory effects. We solve this equation by converting it into a system of ordinary differential equations with the use of a pole expansion of the Fermi-Dirac function. When the applied voltage exceeds the critical value, the superconductor switches to the normal state. We find that when the voltage is decreased from the normal phase, the system relaxes to a Larkin-Ovchinnikov (LO)-type inhomogeneous superconducting state, even in the absence of a magnetic Zeeman field. We point out that the emergence of the LO-type state can be attributed to the nonequilibrium energy distribution of electrons due to the bias voltage. We also point out that the system exhibits bistability, which leads to hysteresis in the voltage-current characteristic of the N-S-N junction.
\end{abstract}
\maketitle
\section{Introduction}
\par 
In condensed-matter physics, nonequilibrium superconductivity has attracted much attention both experimentally and theoretically \cite{GrayBook, Pals1982, Chang1977, Chang1978, TinkhamBook, Testardi1971, Parker1972, Hu1974, Wyatt1966, Dayem1967, Ivlev1973, Sai-Halasz1974, Shuller1976, Chi1981, Sobolewski1986, Bluzer1991, Shi1993, Carr2000, Lucignano2004, Lobo2005, Kusar2008, Beck2013, Demsar2020, Kommers1977, Pals1979, Dynes1972, Tredwell1975, Miller1982, Miller1985, Ginsberg1962, Rothwarf1967, Clarke1972, Tinkham1972, Schmid1975, Scalapino1977, Schmid1977, Schon1977, Hida1978, Gray1978, Eckern1979, Sugahara1979, Heng1981, Tikhonov2018, Gorkov1969, Eliashberg1970, Ivlev1971, Tikhonov2020, Zolochevskii2013, Visser2014, Lara2015, Fuchs1977, Akoh1982,  Arutyunov2011}. It can be realized by exposing a superconductor to external disturbances, such as laser irradiation \cite{Testardi1971, Parker1972, Hu1974, Sai-Halasz1974, Shuller1976, Chi1981, Sobolewski1986, Bluzer1991, Shi1993, Carr2000, Lucignano2004, Lobo2005, Kusar2008, Beck2013, Demsar2020}, microwave irradiation \cite{Wyatt1966, Dayem1967, Ivlev1973, Kommers1977, Pals1979, Visser2014, Lara2015}, phonon injection \cite{Dynes1972, Tredwell1975, Miller1982, Miller1985}, as well as quasiparticle injection \cite{Ginsberg1962, Rothwarf1967, Clarke1972, Fuchs1977, Akoh1982,  Arutyunov2011}. These disturbances cause a deviation of the quasiparticle energy distribution from the equilibrium one, which can be symbolically written as \cite{Pals1982, TinkhamBook}
\begin{equation}
f_{\rm neq}(\omega) = f(\omega) +\Delta f(\omega).
\end{equation}
Here, $f(\omega)$ is the Fermi-Dirac function, and $\Delta f (\omega)$ represents the deviation from the equilibrium distribution. The deviation $\Delta f(\omega)$ leads to various interesting phenomena that cannot be examined as far as we deal with the thermal equilibrium case, such as charge imbalance \cite{Tinkham1972, Schmid1975}, enhancement of superconductivity \cite{Wyatt1966, Dayem1967, Ivlev1973, Kommers1977, Pals1979, Tredwell1975, Gorkov1969, Eliashberg1970, Ivlev1971, Tikhonov2018, Tikhonov2020, Zolochevskii2013, Visser2014, Lara2015}, as well as spatially or temporally inhomogeneous superconducting states \cite{Scalapino1977, Schmid1977, Schon1977, Hida1978, Gray1978, Eckern1979, Sugahara1979, Heng1981}.
\par
In this paper, we revisit nonequilibrium superconductivity realized in a normal metal-superconductor-normal metal (N-S-N) junction \cite{Keizer2006, Moor2009, Snyman2009, Catelani2010, Vercruyssen2012, Serbyn2013, Bobkova2013, Ouassou2018, Seja2021}. When a bias voltage is applied between the normal-metal leads, the quasiparticles are injected into and extracted from the superconductor, which brings the superconductor out of equilibrium. In this sense, a voltage-biased N-S-N junction can be viewed as a driven-dissipative system, where losses of particles and energy in the main system (superconductor) are compensated by environmental systems (normal-metal leads).
\par
In driven-dissipative systems, spatio-temporal pattern formation is commonly found, when the system is driven far from thermal equilibrium by continuous driving in the presence of dissipation \cite{Cross1993, Hoyle2006, Cross2009}. We can thus expect the emergence of a spatially or temporally inhomogeneous superconducting state in a voltage-biased N-S-N junction. Indeed, a N-S-N junction composed of a superconducting wire (the transverse lateral dimension of which is much less than the superconducting coherence length $\xi$) is known to exhibit a time-periodic superconducting state when a constant bias voltage is applied between the normal-metal leads \cite{Langer1967, Skocpol1974, Kramer1977, Ivlev1984, Bezryadin2000, Vodolazov2003, Yerin2013, Yerin2013_2}. The time-periodic superconducting state is associated with the appearance of phase-slip centers (PSCs), at which the superconducting order parameter vanishes.
\par
\begin{figure}[t]
\centering
\includegraphics[width=8.6cm]{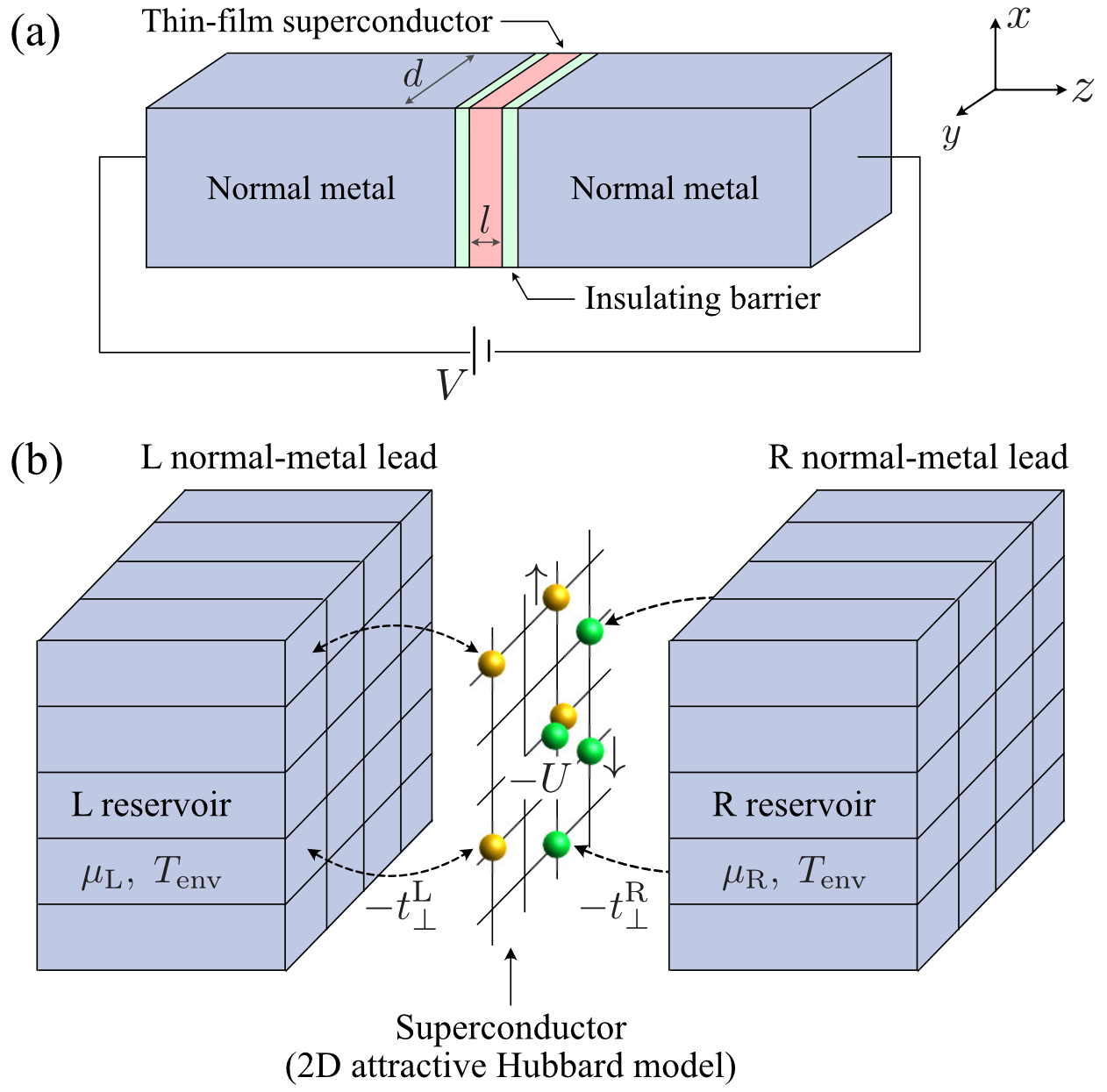}
\caption{(a) The device under consideration. A thin-film ($l\lesssim \xi$ and $d \gg \xi$) superconductor is sandwiched between left (L) and right (R) normal-metal leads. The superconductor is driven out of equilibrium by the bias voltage $V$ between the normal-metal leads. (b) Schematic description of our model. The thin-film superconductor (main system) is described by the two-dimensional attractive Hubbard model. The coupling between the superconductor and the $\alpha$ ($={\rm L}, {\rm R}$) normal-metal lead is modeled by assuming each site of the system is connected to an independent free-fermion $\alpha$ reservoir, which is equilibrated with the chemical potential $\mu_{\alpha}$ and temperature $T_{{\rm env}}$. Electrons are injected into (extracted from) the superconductor from (to) the reservoirs. Moreover, $-t^{\alpha={\rm L}, {\rm R}}_\perp$ describes the hopping amplitude between the superconductor and the $\alpha$ reservoir, which can be controlled by changing the insulating barrier strength inserted between the superconductor and the $\alpha$ normal-metal lead.
}
\label{Fig.model}
\end{figure}
\par
In this paper, we consider a N-S-N junction composed of a thin-film superconductor that is sandwiched by two thick normal metals, as schematically illustrated in Fig.~\ref{Fig.model}(a). We assume that the superconductor in the in-plane direction is much wider than the superconducting coherence length ($d\gg \xi$), but its thickness is small ($l\lesssim\xi$), and the spatial dependence of the superconducting order parameter along the $z$ direction is negligible. In such a N-S-N junction, the time-periodic state associated with the appearance of PSCs is not realized because the cross section (thickness) of the superconductor is too large (small) for PSCs to appear. 
However, we cannot rule out the possibility of a spatially or temporally inhomogeneous superconducting state induced by other nonequilibrium mechanisms than the phase slip.
\par
To explore this possibility, we need to understand the space-time evolution of the superconducting order parameter under the bias voltage. For this purpose, the phenomenological time-dependent Ginzburg-Landau (TDGL) equation has been frequently used \cite{Kramer1977, Ivlev1984, Vodolazov2003, Yerin2013, Yerin2013_2}. Although the TDGL theory is simple and intuitive, it can only be justified in the vicinity of the critical temperature \cite{KopninBook, RammerBook}. Furthermore, the TDGL theory is not sufficient to fully understand the effects of the normal-metal leads on superconductivity. Within the TDGL theory, the effects of the normal-metal leads are often modeled by imposing plausible boundary conditions on the pair wave function \cite{Kramer1977, Ivlev1984, Vodolazov2003, Yerin2013, Yerin2013_2}. However, this phenomenological approach fails to account for the nonequilibrium energy distribution function $f_{\rm neq}(\omega)$ of the electrons in the superconductor. The electrons in the voltage-driven superconductor in Fig.~\ref{Fig.model}(a) obey the nonequilibrium energy distribution function (which will be derived in Sec.~\ref{sec.NESS})
\begin{equation}
f_{\rm neq}(\omega) = 
\frac{\gamma_{\rm L}f_{\rm L}(\omega) +\gamma_{\rm R} f_{\rm R}(\omega)}{\gamma_{\rm L}+\gamma_{\rm R}},
\label{eq.fneq}
\end{equation}
reflecting the different Fermi-Dirac functions $f_\alpha(\omega)$ in the left ($\alpha={\rm L}$) and right ($\alpha={\rm R}$) normal-metal leads due to the bias voltage \cite{Catelani2010}. Here, $\gamma_{\alpha={\rm L}, {\rm R}}$ is the coupling strength between the $\alpha$ normal-metal lead and the superconductor, which will be defined precisely later on. Although the nonequilibrium energy distribution function $f_{\rm neq}(\omega)$ in Eq.~\eqref{eq.fneq} can have a significant impact on superconductivity, the TDGL theory cannot capture this.
\par
In this paper, using the nonequilibrium Green's function technique \cite{RammerBook, ZagoskinBook,StefanucciBook}, we derive a quantum kinetic equation for nonequilibrium superconductivity in the N-S-N junction, to overcome the shortcomings of the above-mentioned TDGL theory. Solving this equation, we determine the superconducting order parameter in the voltage-driven nonequilibrium superconductor. We clarify that the nonequilibrium energy distribution $f_{\rm neq}(\omega)$ induces a  steady but {\it spatially inhomogeneous} nonequilibrium superconductivity. We also show that the N-S-N junction system exhibits bistability, leading to hysteresis in the voltage-current characteristic of the junction.
\par
We note that the derived quantum kinetic equation is a non-Markovian integro-differential equation with memory effects. Although the numerical solution of such an equation is usually a computationally very expensive problem, we show that one can effectively reduce the associated numerical costs by utilizing the auxiliary-mode expansion technique, originally developed to study the time-dependent electron transport in a normal-metal device \cite{Croy2009, Croy2011, Croy2012, Popescu2018, Lehmann2018, Tuovinen2023}.
\par
We comment on the connection between previous work \cite{Moor2009, Bobkova2013, Snyman2009} and the present study: In Refs.~\cite{Moor2009, Bobkova2013}, the possibility of a spatially inhomogeneous superconducting state in a voltage-driven superconductor was discussed based on the quasiclassical Green's function technique. However, these papers do not solve the quantum kinetic equation to study the time evolution of the system. Thus, the possibility of a temporally inhomogeneous superconducting state is neglected. Besides, the stability of the spatially inhomogeneous superconducting state and the bistability are not explicitly discussed. On the other hand, Ref.~\cite{Snyman2009} examines the dynamics of the voltage-biased superconductor, to point out the possibility of a bistability. However, Ref.~\cite{Snyman2009} deals only with the spatially homogeneous case. Our work complements these previous studies, to give the complete phase diagram of a voltage-biased superconductor.
\par
This paper is organized as follows. In Sec.~\ref{sec.formalism}, we derive the quantum kinetic equation for nonequilibrium superconductivity in the N-S-N junction. By solving this equation, we draw the nonequilibrium phase diagram of this system in Sec.~\ref{sec.result}. Throughout this paper, we use units such that $\hbar=k_{\rm B}=1$, and the volumes of the reservoirs are taken to be unity, for simplicity.
\par
\par
\section{Formalism \label{sec.formalism}}
\par 
\subsection{Model Hamiltonian}
We consider a thin-film superconductor sandwiched between normal-metal leads, as illustrated in Fig.~\ref{Fig.model}(a). To model this N-S-N junction, we consider the Hamiltonian,
\begin{equation}
H= H_{\rm SC} + H_{\rm lead} + H_{\rm mix}.
\label{eq.Htot*}
\end{equation}
Here, the superconductor (which is referred to as the main system in what follows) is described by the two-dimensional attractive Hubbard model on a square lattice of $N=L_x \times L_y$ sites with the periodic boundary conditions, described by
\begin{align}
&
H_{\rm SC} =H_0 + H_{\rm int},\label{eq.Hsc}\\
&
H_0= -t_{\parallel} \sum_{\sigma=\up,\down} \sum_{(j,k)} \big[c^\dagger_{j, \sigma} c_{k, \sigma} +{\rm H.c.} \big] -\mu_{\rm sys} \sum_{\sigma=\up, \down} \sum_{j=1}^N n_{j,\sigma}, \label{eq.H0*}\\
&
H_{\rm int}= -U \sum_{j=1}^{N}n_{j,\up} n_{j,\down},
\label{eq.Hint}
\end{align}
where $c_{j, \sigma}$ is the annihilation operator of an electron with spin $\sigma=\up, \down$ at the $j$th lattice site ($j=1,\cdots, N$), and $n_{j,\sigma}=c^\dagger_{j, \sigma}c_{j, \sigma}$ is the number operator. In Eq.~\eqref{eq.H0*}, $-t_\parallel$ is the nearest-neighbor hopping amplitude, $\mu_{\rm sys}$ is the chemical potential of the main system, and the summation $(j, k)$ is taken over the nearest-neighbor lattice sites. The Hubbard-type on-site interaction is described by $H_{\rm int}$ in Eq.~\eqref{eq.Hint}, where $-U$ $(<0)$ is the strength of the attractive pairing interaction.
\par
\begin{figure}[t]
\centering
\includegraphics[width=8.6cm]{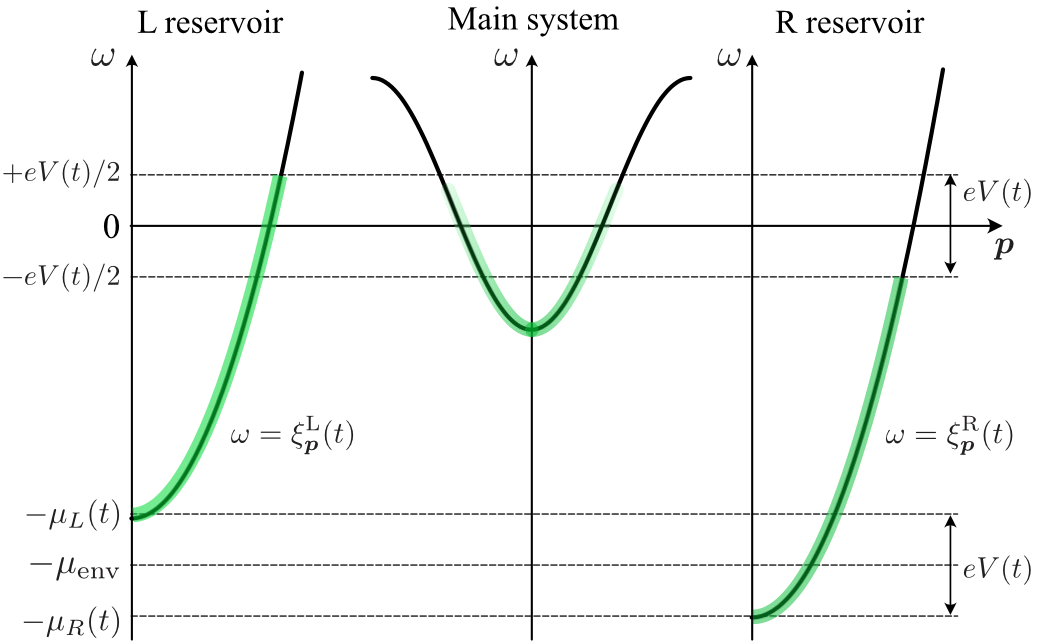}
\caption{Schematic energy band structure of our model. The energy is commonly measured from the average of the Fermi energy levels of the left and right reservoirs. In the left (right) reservoir at $T_{\rm env}=0$, the energy band $\xi^{{\rm L} ({\rm R})}_{\bm{p}}(t)$ is filled up to $\pm eV(t)/2$, respectively. The chemical potential difference $\mu_{\rm R}(t)-\mu_{\rm L}(t)$ between the left and right reservoirs equals the applied bias voltage $e V(t)$ between the normal-metal leads.}
\label{Fig.energy.band}
\end{figure}
\par
The left ($\alpha={\rm L}$) and right ($\alpha={\rm R}$) normal-metal leads are assumed to be free-electron gases in a thermal equilibrium state, described by $H_{\rm lead}$ in Eq.~\eqref{eq.Htot*}, having the form,
\begin{equation}
H_{\rm lead} = \sum_{\alpha={\rm L}, {\rm R}} \sum_{\sigma=\up, \down} \sum_{j=1}^N  \sum_{\bm{p}} \xi_{\bm{p}}^\alpha(t) d^{\alpha \dagger}_{j, \bm{p}, \sigma} d^{\alpha}_{j, \bm{p}, \sigma},
\label{eq.Hlead}
\end{equation}
where $d^{\alpha \dagger}_{j, \bm{p},\sigma}$ creates an electron with the kinetic energy $\xi^\alpha_{\bm{p}}=\bm{p}^2/2m -\mu_\alpha(t)$ in the $j$th $\alpha$ reservoir. We model the coupling between the main system and the $\alpha$ normal-metal lead by assuming each site of the system is connected to an independent free-fermion reservoir (which we call the $\alpha$ reservoir) \cite{Aoki2014}, as schematically illustrated in Fig.~\ref{Fig.model}(b). All the $\alpha$ reservoirs are assumed to be always equilibrated with the chemical potential $\mu_{\alpha}(t)$ and the temperature $T_{\rm env}$ \cite{note.T}. The difference $\mu_{\rm R}(t)-\mu_{\rm L}(t)$ of the chemical potentials between the left and right reservoirs equals the applied bias voltage $eV(t)$ between the normal-metal leads.  As schematically shown in Fig.~\ref{Fig.energy.band}, we parametrize $\mu_\alpha(t)$ as
\begin{align}
\mu_{\rm L(R)}(t) 
&=
\mu_{\rm env} \mp \frac{e}{2}V(t)
\notag\\
&= \mu_{\rm env} \mp \frac{e}{2}\big[ V_0 +\Delta V(t) \Theta(t) \big],	
\label{eq.muLR}
\end{align}
where $\mu_{\rm env}$ is the average chemical potential and $\Theta(t)$ is the step function. We assume that the system is in a nonequilibrium steady state (NESS) under a constant voltage $V_0$ at $t=0$. (We will verify this assumption later by investigating the time evolution of the system.) For $t>0$, the system is driven by a time-dependent voltage $V_0 +\Delta V(t)$.
\par
The couplings between the superconductor and the normal-metal leads are described by
\begin{equation}
H_{\rm mix} = -\sum_{\alpha={\rm L}, {\rm R}} \sum_{\sigma=\up, \down} \sum_{j=1}^N  \sum_{\bm{p}}\big[t_\perp^\alpha d^{\alpha \dagger}_{j, \bm{p}, \sigma} c_{j,\sigma} +{\rm H.c.}\big].
\end{equation}
Here, $-t_\perp^{\alpha}$ is the hopping amplitude between the superconductor and the $\alpha$ reservoir, which can be tuned by adjusting the insulating barrier strength inserted between the $\alpha$ normal-metal lead and the superconductor. In this paper, for simplicity, we consider the symmetric hopping amplitude $t_\perp^{\rm L}=t_\perp^{\rm R} \equiv t_\perp$.
\par
\subsection{Nonequilibrium Nambu Green's function}
\par
To study the superconducting state out of equilibrium, we conveniently introduce a $2N\times 2N$ matrix nonequilibrium Nambu Green's functions, given by
\begin{equation}
\hat{\bm{G}}^{\rm x={\rm r}, {\rm a}, <}(t,t')=
\begin{pmatrix}
\bm{G}^{\rm x}_{11}(t,t') &  \cdots & \bm{G}^{\rm x}_{1N}(t,t') \\
\vdots & \ddots & \vdots \\
\bm{G}^{\rm x}_{N1}(t,t') & \cdots & \bm{G}^{\rm x}_{NN}(t,t')
\end{pmatrix}_{2N\times 2N},
\end{equation}
where
\begin{align}
\bm{G}^{\rm r}_{jk}(t,t')
&=
-i\Theta(t-t')
\notag\\
&\hspace{0.2cm}\times
\begin{pmatrix}
\braket{[c_{j,\up}(t),  c^\dagger_{k,\up}(t')]_+} &  
\braket{[c_{j,\up}(t),  c_{k,\down}(t')]_+} \\[4pt]
\braket{[c^\dagger_{j,\down}(t),  c^\dagger_{k,\up}(t')]_+} & 
\braket{[c^\dagger_{j,\down}(t),  c_{k,\down}(t')]_+} 
\end{pmatrix}
\notag\\
&=
\big[\bm{G}^{\rm a}_{jk}(t', t)\big]^\dagger
\label{eq.Gra}
,\\[4pt]
\bm{G}^<_{jk}(t,t') 
&=
\begin{pmatrix}
i\braket{c^\dagger_{k, \up}(t') c_{j, \up}(t)} & 
i\braket{c_{k, \down}(t') c_{j, \up}(t)} \\[4pt]
i\braket{c^\dagger_{k, \up}(t') c^\dagger_{j, \down}(t)} &
i\braket{c_{k, \down}(t') c^\dagger_{j, \down}(t)}
\end{pmatrix},
\label{eq.G<}
\end{align}
with $[A, B]_\pm = AB \pm BA$. In Eqs.~\eqref{eq.Gra} and \eqref{eq.G<}, $\bm{G}^{\rm r}_{jk}$, $\bm{G}^{\rm a}_{jk}$, and $\bm{G}^{<}_{jk}$ are, respectively, the retarded, advanced, and lesser $2\times 2$ matrix Green’s functions, whose elements are given by
\begin{equation}
\bm{G}^{{\rm x}={\rm r}, {\rm a}, <}_{jk}(t,t')=
\begin{pmatrix}
G^{\rm x}_{jk}(t, t')_{11} & G^{\rm x}_{jk}(t, t')_{12} \\[4pt]
G^{\rm x}_{jk}(t, t')_{21} & G^{\rm x}_{jk}(t, t')_{22}
\end{pmatrix}.
\end{equation}
\par
In the nonequilibrium Green's function scheme, the effects of the pairing interaction, as well as the system-lead couplings, can be summarized by the $2N\times 2N$ matrix self-energy correction
\begin{align}
\hat{\bm{\Sigma}}^{\rm x={\rm r}, {\rm a}, <}(t,t')
&=
\begin{pmatrix}
\bm{\Sigma}^{\rm x}_{11}(t,t') &  \cdots & \bm{\Sigma}^{\rm x}_{1N}(t,t') \\
\vdots & \ddots & \vdots \\
\bm{\Sigma}^{\rm x}_{N1}(t,t') & \cdots & \bm{\Sigma}^{\rm x}_{NN}(t,t')
\end{pmatrix}_{2N\times 2N}
\notag\\
&=
\hat{\bm{\Sigma}}^{\rm x}_{\rm int}(t,t')+
\hat{\bm{\Sigma}}^{\rm x}_{\rm lead}(t,t'),
\label{eq.self}
\end{align}
which appears in the Keldysh-Dyson equations \cite{RammerBook, ZagoskinBook,StefanucciBook, Jauho1994},
\begin{align}
&
\hat{\bm{G}}^{\rm r(a)}(t,t')=
\hat{\bm{G}}^{\rm r(a)}_0(t,t') 
\notag\\[4pt]
&\hspace{0.4cm}+
\int_{-\infty}^\infty dt_1 \int_{-\infty}^\infty dt_2
\hat{\bm{G}}^{\rm r(a)}_0(t, t_1) \hat{\bm{\Sigma}}^{\rm r(a)}(t_1, t_2) \hat{\bm{G}}^{\rm r(a)}(t_2, t')
\label{eq.Dyson.GR},\\
&
\hat{\bm{G}}^{<}(t,t')=
\hat{\bm{G}}^{<}_{\rm iso}(t,t')
\notag\\[4pt]
&\hspace{0.5cm}+
\int_{-\infty}^\infty dt_1 \int_{-\infty}^\infty dt_2\hat{\bm{G}}^{\rm r} (t, t_1)\hat{\bm{\Sigma}}^< (t_1, t_2)\hat{\bm{G}}^{\rm a}(t_2, t'),
\label{eq.Dyson.GL.with.ini}
\end{align}
with
\begin{align}
&
\hat{\bm{G}}^{<}_{\rm iso}(t,t')=
\int_{-\infty}^\infty dt_1 \int_{-\infty}^\infty dt_2
\notag\\
&\hspace{0.4cm}\times
\left[
\delta(t -t_1)
+ \int_{-\infty}^\infty dt_3
\hat{\bm{G}}^{\rm r}(t, t_3) \hat{\bm{\Sigma}}^{\rm r}(t_3, t_1)
\right]
\hat{\bm{G}}_0^<(t_1,t_2) 
\notag\\
&\hspace{0.4cm}\times
\left[
\delta(t_2 -t')+ \int_{-\infty}^\infty dt_3
\hat{\bm{\Sigma}}^{\rm a}(t_2, t_3) \hat{\bm{G}}^{\rm a}(t_3, t')
\right].
\label{eq.GL.ini**}
\end{align}
Here, $\hat{\bm{G}}^{\rm r(a)}_0$ and $\hat{\bm{G}}_0^<$ are the Green’s functions of the isolated main system in the thermal equilibrium state {\it before} the system-lead couplings, as well as the pairing interaction, are switched on. We note that although $\hat{\bm{G}}_0^<$ depends on the temperature $T_{\rm iso}$ in the isolated main system, we will later find that $\hat{\bm{G}}_{\rm iso}^<$ in Eq.~\eqref{eq.GL.ini**}, which involves $\hat{\bm{G}}_0^<$, vanishes within our formalism and the results do not depend on $T_{\rm iso}$. (Instead,  the results only depend on the temperature $T_{\rm env}$ in the normal-metal leads.)
\par
In Eq.~\eqref{eq.self}, $\hat{\bm{\Sigma}}^{\rm x}_{\rm int}$ and $\hat{\bm{\Sigma}}^{\rm x}_{\rm lead}$ describe the effects of the pairing interaction and the system-lead couplings, respectively. In the mean-field BCS approximation, $\hat{\bm{\Sigma}}^{\rm x}_{\rm int}$ is given by \cite{Yamaguchi2012, Hanai2017, Kawamura2022},
\begin{align}
\hat{\bm{\Sigma}}^{\rm r}_{\rm int}(t, t')
&=
\hat{\bm{\Sigma}}^{\rm a}_{\rm int}(t, t')
=
-\hat{\bm{\Delta}}(t)\delta(t-t'),
\label{eq.selfR.int}
\\
\hat{\bm{\Sigma}}^{<}_{\rm int}(t, t')&=0,
\label{eq.selfL.int}
\end{align}
where $\hat{\Delta}(t)$ denotes the $2N\times 2N$ matrix superconducting order parameter, having the form,  
\begin{equation}
\hat{\bm{\Delta}}(t)=
\begin{pmatrix}
\bm{\Delta}_1(t) &  \\
 & \ddots \\
&& \bm{\Delta}_N(t)
\end{pmatrix}_{2N\times 2N}
\end{equation}
with
\begin{equation}
\bm{\Delta}_j(t) =
\begin{pmatrix}
0 & \Delta_j(t)  \\[4pt]
\Delta^*_j(t) & 0
\end{pmatrix}.
\label{eq.OP.Nambu}
\end{equation}
In Eq.~\eqref{eq.OP.Nambu}, 
\begin{equation}
\Delta_j(t) 
= -iU G^<_{jj}(t, t)_{12}
\label{eq.OP}
\end{equation}
is the local superconducting order parameter of the $j$th lattice site.
\par
We deal with the self-energy correction $\hat{\bm{\Sigma}}^{\rm x}_{\rm lead}$ within the second-order Born approximation with respect to the hopping amplitude $t_\perp$, which gives (For the derivation, see Appendix~\ref{sec.App.self}.)
\begin{align}
\hat{\bm{\Sigma}}^{\rm r}_{\rm lead}(t, t')
&=
\big[\hat{\bm{\Sigma}}^{\rm a}_{\rm lead}(t', t)\big]^\dagger=
-2i\gamma \delta(t-t') \hat{\bm{1}}
\label{eq.selfR.env}
,\\[4pt]
\hat{\bm{\Sigma}}^<_{\rm lead}(t, t')
&=
2i\gamma \sum_{\eta=\pm}
\exp\left(-i \eta \int_{t'}^t dt_1\hspace{0.1cm} \frac{e\Delta V(t_1)}{2}\right)
\notag\\
&\hspace{0.9cm}\times
\int_{-\infty}^\infty \frac{d\omega}{2\pi}e^{-i\omega(t-t')} f\left(\omega -\eta \frac{eV_{0}}{2}\right) \hat{\bm{1}}.
\label{eq.selfL.env}
\end{align}
Here, $\hat{\bm{1}}$ is the $2N\times 2N$ unit matrix, 
\begin{equation}
f(\omega) = \frac{1}{e^{\omega/T_{\rm env}}+1}
\end{equation}
is the Fermi-Dirac distribution function in the reservoirs, and
\begin{equation}
\gamma=\pi \rho |t_\perp|^2,
\label{eq.gamma}
\end{equation}
with $\rho_\alpha(\omega)\equiv \rho$ being the single-particle density of state in the $\alpha$ reservoirs. In the following, we use $\gamma$ as a parameter to characterize the system-lead coupling strength. We note that in deriving $\hat{\bm{\Sigma}}^{\rm x}_{\rm lead}$, we assume that $\rho_\alpha(\omega)$ in the $\alpha$ normal-metal lead are unperturbed by the proximity effect because the system-lead couplings are sufficiently weak \cite{Moor2009, Snyman2009}. Under this assumption, we can ignore the $\omega$ dependence of $\rho_\alpha(\omega)$ around the Fermi levels, which is called the wide-band approximation in the literature \cite{StefanucciBook, Aoki2014}. We also note that we have ignored the real part of the self-energy $\hat{\bm{\Sigma}}^{\rm r(a)}_{\rm lead}$ in Eq.~\eqref{eq.selfR.env}, because it only gives a constant energy shift, which can be absorbed into the chemical potential $\mu_{\rm sys}$ of the main system \cite{StefanucciBook, Aoki2014}.
\par
Due to the wide-band approximation, the retarded (advanced) self-energy $\hat{\bm{\Sigma}}^{\rm r(a)}_{\rm lead}(t, t')$ in Eq.~\eqref{eq.selfR.env} becomes local in time, that is, it has non-zero contribution only when $t=t'$. On the other hand, the lesser self-energy $\hat{\bm{\Sigma}}^{<}_{\rm lead}(t, t')$ in Eq.~\eqref{eq.selfL.env} is nonlocal in time, because we do not ignore the $\omega$ dependence of the energy distribution function $f(\omega)$ in the $\alpha$ reservoir.
\par
As shown in Appendix~\ref{eq.sec.app.vanish.Gini}, substituting the self-energy corrections in Eqs.~\eqref{eq.selfR.int}, \eqref{eq.selfL.int}, \eqref{eq.selfR.env}, and \eqref{eq.selfL.env} into the Keldysh-Dyson equations \eqref{eq.Dyson.GR} and \eqref{eq.Dyson.GL.with.ini}, we find $\hat{\bm{G}}^{<}_{\rm iso}(t,t')=0$, which simplifies Eq.~\eqref{eq.Dyson.GL.with.ini} as \cite{Jauho1994}
\begin{equation}
\hat{\bm{G}}^{<}(t,t')=
\int_{-\infty}^\infty dt_1 \int_{-\infty}^\infty dt_2\hat{\bm{G}}^{\rm r} (t, t_1)\hat{\bm{\Sigma}}^< (t_1, t_2)\hat{\bm{G}}^{\rm a}(t_2, t').
\label{eq.Dyson.GL}
\end{equation}
We note that the vanishing of $\hat{\bm{G}}^{<}_{\rm iso}(t,t')$ physically means that the memory of the thermal equilibrium state in the isolated main system is wiped out by the couplings with the normal-metal leads.
\par
\subsection{Nonequilibrium steady state at $t=0$ \label{sec.NESS}}
\par
We next explain how to obtain the nonequilibrium superconducting steady state at $t=0$ under the constant voltage $V_0$. When the system is in a NESS, the Green's functions $\hat{\bm{G}}^{{\rm x}={\rm r}, {\rm a}, <}(t, t')$, as well as the self-energy corrections $
\hat{\bm{\Sigma}}^{\rm x}(t, t')$, depend only on the relative time $t-t'$. This simplifies the Keldysh-Dyson equations~\eqref{eq.Dyson.GR} and \eqref{eq.Dyson.GL} as
\begin{align}
&
\hat{\bm{G}}^{\rm r(a)}_{\rm NESS}(\omega)=
\hat{\bm{G}}^{\rm r(a)}_0(\omega)+
\hat{\bm{G}}^{\rm r(a)}_0(\omega) 
\hat{\bm{\Sigma}}^{\rm r(a)}(\omega) 
\hat{\bm{G}}^{\rm r(a)}_{\rm NESS}(\omega)
\label{eq.Dyson.GR.NESS},\\
&
\hat{\bm{G}}^{<}_{\rm NESS}(\omega)=
\hat{\bm{G}}^{\rm r}_{\rm NESS}(\omega)
\hat{\bm{\Sigma}}^<(\omega)
\hat{\bm{G}}^{\rm a}_{\rm NESS}(\omega),
\label{eq.Dyson.GL.NESS}
\end{align}
where
\begin{align}
&
\hat{\bm{G}}^{{\rm x}}_{\rm NESS}(\omega)=
\int_{-\infty}^\infty d(t-t') e^{i\omega (t-t')}\hat{\bm{G}}^{\rm x}_{\rm NESS}(t-t')
,\\
&
\hat{\bm{\Sigma}}^{{\rm x}}(\omega)=
\int_{-\infty}^\infty d(t-t') e^{i\omega (t-t')}\hat{\bm{\Sigma}}^{\rm x}(t-t').
\end{align}
In frequency space, Eqs.~\eqref{eq.selfR.int} and \eqref{eq.selfL.int} have the forms
\begin{align}
& \hat{\bm{\Sigma}}^{\rm r(a)}_{\rm int}(\omega)= -\hat{\bm{\Delta}}(t=0)
\label{eq.self.R.int.NESS}
,\\
& \hat{\bm{\Sigma}}^{<}_{\rm int}(\omega)=0.
\label{eq.self.L.int.NESS}
\end{align}	
In the same manner, the self-energy $\hat{\bm{\Sigma}}^{\rm x}_{\rm lead}(\omega)$ is given by
\begin{align}
&
\hat{\bm{\Sigma}}^{\rm r(a)}_{\rm lead}(\omega)= \mp 2i\gamma \hat{\bm{1}}
\label{eq.self.R.lead.NESS}
,\\
&
\hat{\bm{\Sigma}}^{<}_{\rm int}(\omega) = 2i\gamma \left[f\left(\omega -\frac{eV_0}{2}\right) +f\left(\omega +\frac{eV_0}{2}\right)\right] \hat{\bm{1}}.
\label{eq.self.L.lead.NESS}
\end{align}
Substituting Eqs.~\eqref{eq.self.R.int.NESS}-\eqref{eq.self.L.lead.NESS} into the Dyson equation~\eqref{eq.Dyson.GR.NESS}, we have
\begin{equation}
\hat{\bm{G}}^{\rm r(a)}_{\rm NESS}(\omega)= \frac{1}{\omega \pm 2i\gamma -\hat{\bm{\m{H}}}_{\rm BdG}},
\label{eq.GR.NESS}
\end{equation}
where
\begin{equation}
\hat{\bm{\m{H}}}_{\rm BdG} =\hat{\bm{\m{H}}}_0 -\hat{\bm{\Delta}}(0),
\end{equation}
with $\hat{\bm{\m{H}}}_0$ being the matrix representation of the Hamiltonian $H_0$ in Eq.~\eqref{eq.H0*}, given by
\begin{equation}
H_0 =\hat{\bm{\Psi}}^\dagger \hat{\bm{\m{H}}}_0 \hat{\bm{\Psi}}.
\label{eq.H0}
\end{equation}
In Eq,~\eqref{eq.H0},
\begin{equation}
\hat{\bm{\Psi}}^\dagger = \big(c^\dagger_{1,\up}, c_{1,\down}, \cdots, c^\dagger_{N,\up} c_{N,\down} \big).
\label{eq.Nambu.field}
\end{equation}
is the $2N$-component Nambu field.
\par 
The BCS mean-field Hamiltonian $\hat{\bm{\m{H}}}_{\rm BdG}$ can be diagonalized by the Bogoliubov transformation,
\begin{equation}
\hat{\bm{W}}^\dagger \hat{\bm{\m{H}}}_{\rm BdG} \hat{\bm{W}} =
\begin{pmatrix}
E_1 &  \\
 & \ddots\\
&& E_{2N}
\end{pmatrix}_{2N\times 2N}.
\label{eq.def.Bogo}
\end{equation}
Here, $\hat{\bm{W}}$ is a unitary matrix and $E_{j=1,\cdots,2N}$ are eigenvalues of $\hat{\bm{\m{H}}}_{\rm BdG}$. The retarded (advanced) Green's function in Eq.~\eqref{eq.GR.NESS} can also be diagonalized by using $\hat{\bm{W}}$ as
\begin{widetext}
\begin{equation}
\hat{\bm{\m{G}}}^{\rm r(a)}_{\rm NESS}(\omega)
\equiv	
\hat{\bm{W}}^\dagger \hat{\bm{G}}^{\rm r(a)}_{\rm NESS}(\omega) \hat{\bm{W}} 
=
\begin{pmatrix}
[\omega \pm 2i\gamma -E_1]^{-1} &  \\
 & \ddots \\
&& [\omega \pm 2i\gamma -E_{2N}]^{-1}
\end{pmatrix}_{2N\times 2N}.
\label{eq.Bogliubov.GR}
\end{equation}
From Eqs.~\eqref{eq.Dyson.GL.NESS}, \eqref{eq.self.L.int.NESS}, \eqref{eq.self.L.lead.NESS}, and \eqref{eq.Bogliubov.GR}, we obtain the lesser component $\hat{\bm{\m{G}}}^{<}_{\rm NESS}$ of the Green's function as
\begin{equation}
\hat{\bm{\m{G}}}^{<}_{\rm NESS}(\omega)	
\equiv 
\hat{\bm{W}}^\dagger \hat{\bm{G}}^{<}_{\rm NESS}(\omega) \hat{\bm{W}}
=2i\gamma
\begin{pmatrix}
\frac{f\left(\omega -eV_0/2\right) +f\left(\omega +eV_0/2\right)}{[\omega-E_1]^2 +4\gamma^2}&  \\
 & \ddots \\
&& \frac{f\left(\omega -eV_0/2\right) +f\left(\omega +eV_0/2\right)}{[\omega-E_{2N}]^2 +4\gamma^2}
\end{pmatrix}_{2N\times 2N}.
\label{eq.Boboliubov.GL}
\end{equation}
\par
The local superconducting order parameter $\Delta_j(0)$ is self-consistently determined from Eq.~\eqref{eq.OP} as
\begin{equation}
\Delta_j(0) 
=
\int_{-\infty}^\infty \frac{d\omega}{2\pi}
\hat{\bm{G}}^<_{\rm NESS}(\omega)_{2j-1, 2j}	
=
\big[\hat{\bm{W}} \hat{\bm{\Lambda}} \hat{\bm{W}}^\dagger \big]_{2j-1, 2j},
\label{eq.gap.NESS}
\end{equation}
where
\begin{equation}
\hat{\bm{\Lambda}}=
\sum_{\eta=\pm} \hat{\bm{\Lambda}}_\eta =
\int_{-\infty}^\infty \frac{d\omega}{2\pi}\hat{\bm{\m{G}}}^{<}_{\rm NESS}(\omega),
\label{eq.def.Lamb}
\end{equation}
with $\hat{\bm{\Lambda}}_\eta={\rm diag}(\Lambda^\eta_1, \cdots, \Lambda^\eta_{2N})$ being a $2N\times 2N$ diagonal matrix. In Eq.~\eqref{eq.def.Lamb}, $\Lambda^\eta_{j=1, \cdots, 2N}$ are given by
\begin{align}
\Lambda^\eta_{j} 
&=
2i\gamma \int_{-\infty}^\infty \frac{d\omega}{2\pi} \frac{f\left(\omega -\eta e V_0/2\right)}{[\omega-E_j]^2 +4\gamma^2}
\notag\\
&= 
\frac{i}{2} 
\Bigg[\frac{1}{e^{(E_j +2i\gamma -\eta e V_0)/2}+1} +
\frac{1}{2\pi i} \left[
\psi\left(\frac{1}{2} -\frac{E_j+2i\gamma -\eta e V_0/2}{2\pi i T_{\rm env}}\right) -
\psi\left(\frac{1}{2} -\frac{E_j-2i\gamma -\eta e V_0/2}{2\pi i T_{\rm env}}\right)
\right]\Bigg].
\label{eq.full.Lamb}
\end{align}
Here, $\psi(z)$ is a complex digamma function. We solve the gap equation \eqref{eq.gap.NESS}, to obtain the superconducting order parameter $\Delta_j(0)$ for a given set ($\gamma, V_0, T_{\rm env}$) of parameters. For this purpose, we employ the restarted Pulay mixing scheme \cite{Pratapa2015, Banerjee2016}, to accelerate the convergence of this self-consistent calculation.
\end{widetext}
\par
We note that substituting Eqs.~\eqref{eq.self.L.int.NESS} and \eqref{eq.self.L.lead.NESS} into the Dyson equation~\eqref{eq.Dyson.GL.NESS}, we have
\begin{align}
\hat{\bm{G}}^{<}_{\rm NESS}(\omega)
=&
 2i\gamma \left[f\left(\omega -\frac{eV_0}{2}\right) +f\left(\omega +\frac{eV_0}{2}\right)\right]
\notag\\
&\hspace{2cm}\times
\hat{\bm{G}}^{\rm r}_{\rm NESS}(\omega)
\hat{\bm{G}}^{\rm a}_{\rm NESS}(\omega)	
\notag\\[4pt]
&=
-\frac{1}{2}\left[f\left(\omega -\frac{eV_0}{2}\right) +f\left(\omega +\frac{eV_0}{2}\right)\right] 
\notag\\
&\hspace{2cm}\times
\big[\hat{\bm{G}}^{\rm r}_{\rm NESS}(\omega) -\hat{\bm{G}}^{\rm a}_{\rm NESS}(\omega)\big]
\notag\\
&\equiv
i f_{\rm neq}(\omega) \hat{\bm{A}}_{\rm NESS}(\omega),
\label{eq.fneq.deriv}
\end{align}
with
\begin{align}
& f_{\rm neq}(\omega) =
\frac{1}{2}\left[f\left(\omega -\frac{eV_0}{2}\right) +f\left(\omega +\frac{eV_0}{2}\right)\right] 
\label{eq.fneq.**}
,\\[4pt]
& \hat{\bm{A}}_{\rm NESS}(\omega) = i\big[\hat{\bm{G}}^{\rm r}_{\rm NESS}(\omega) -\hat{\bm{G}}^{\rm a}_{\rm NESS}(\omega)\big].
\end{align}
In deriving the second line of Eq.~\eqref{eq.fneq.deriv}, we used \cite{JishiBook}
\begin{align}
&
\hat{\bm{G}}^{\rm r}_{\rm NESS}(\omega) -\hat{\bm{G}}^{\rm a}_{\rm NESS}(\omega)
\notag\\
&\hspace{1.2cm}=
\hat{\bm{G}}^{\rm r}_{\rm NESS}(\omega)
\big[ 
\hat{\bm{\Sigma}}^{\rm r}(\omega)-
\hat{\bm{\Sigma}}^{\rm a}(\omega)
\big]
\hat{\bm{G}}^{\rm a}_{\rm NESS}(\omega)
\notag\\
&\hspace{1.2cm}=
-4i\gamma \hat{\bm{G}}^{\rm r}_{\rm NESS}(\omega) \hat{\bm{G}}^{\rm a}_{\rm NESS}(\omega).
\end{align}
Comparing Eq.~\eqref{eq.fneq.deriv} with the thermal equilibrium lesser Green's function, we can interpret $\hat{\bm{A}}_{\rm NESS}(\omega)$ and $f_{\rm neq}(\omega)$ as the single-electron excitation spectra and the nonequilibrium energy distribution, respectively. We note that the nonequilibrium energy distribution $f_{\rm neq}(\omega)$ in Eq.~\eqref{eq.fneq.**} is given in Eq.~\eqref{eq.fneq} where $\gamma_{\rm L}=\gamma_{\rm R}$.
\par
Once the nonequilibrium superconducting steady state is obtained by solving the gap equation \eqref{eq.gap.NESS}, we can evaluate the steady-state charge current $I$ through the N-S-N junction. The charge current $I_{\rm L}(t)$ from the left normal-metal lead to the superconductor is determined from the rate of change in the number of electrons in the left reservoirs \cite{Meir1992, Jauho1994}:
\begin{widetext}
\begin{align}
I_{\rm L}(t) 
&= 
-e \frac{d}{dt} \sum_{\sigma=\up, \down} \sum_{j=1}^N \sum_{\bm{p}} \braket{d^{{\rm L}\dagger}_{j, \bm{p}, \sigma}(t) d^{{\rm L}}_{j, \bm{p}, \sigma}(t)}
\notag\\[4pt]
&=
2e |t_\perp|^2 \sum_{j=1}^N \sum_{\bm{p}} 
\int_{-\infty}^\infty dt_1{\rm Re} 
\Big[ G^{\rm r}_{jj}(t, t_1)_{11} D^<_{\rm L}(\bm{p}, t_1, t)_{11} +
G^<_{jj}(t, t_1)_{11} D^{\rm a}_{\rm L}(\bm{p}, t_1, t)_{11} \Big].
\label{eq.JLt}
\end{align}
Here, $\bm{D}^{{\rm a} (<)}_{\rm L}(\bm{p}, t, t')$ is the non-interacting advanced (lesser) Green's function in the left reservoir, given in Eqs.~\eqref{eq.app.self.lead.a} and \eqref{eq.app.self.lead.<}. From Eq.~\eqref{eq.JLt}, we obtain the steady-state value $I_{\rm L}(t=0)$ as \cite{Meir1992, Jauho1994}
\begin{equation}
I_{\rm L} = 4i\gamma e \sum_{j=1}^N \int_{-\infty}^\infty \frac{d\omega}{2\pi}
\left[f\left(\omega-\frac{eV_0}{2}\right) 
\big[G^{\rm r}_{{\rm NESS}, jj}(\omega)_{11}-G^{\rm a}_{{\rm NESS}, jj}(\omega)_{11}\big] +G^<_{{\rm NESS}, jj}(\omega)_{11}\right].
\label{eq.JL}
\end{equation}
The current $I_{\rm R}$ from the right lead to the superconductor is also given by Eq.~\eqref{eq.JL} where $f(\omega -eV_0/2)$ is replaced by $f(\omega +eV_0/2)$. Since $I \equiv I_{\rm L} = -I_{\rm R}$ in a NESS, we obtain a symmetrical expression $I=[I_{\rm L}-I_{\rm R}]/2$ as
\begin{equation}
I=2i\gamma e \sum_{j=1}^N \int_{-\infty}^\infty\frac{d\omega}{2\pi}
\left[f\left(\omega-\frac{eV_0}{2}\right)  -f\left(\omega+\frac{eV_0}{2}\right) \right] \big[G^{\rm r}_{{\rm NESS}, jj}(\omega)_{11}-G^{\rm a}_{{\rm NESS}, jj}(\omega)_{11}\big].
\label{eq.sym.J}
\end{equation}
\end{widetext}
The $\omega$ integral in Eq.~\eqref{eq.sym.J} can be performed analytically by employing the Bogoliubov transformation in Eq.~\eqref{eq.def.Bogo}. After carrying out the Bogoliubov transformation, we have
\begin{equation}
I =4\gamma e \sum_{j=1}^N \Big[\hat{\bm{W}}\big[\hat{\bm{\Lambda}}_+ -\hat{\bm{\Lambda}}_-\big] \bm{W}^\dagger\Big]_{2j-1, 2j-1},
\end{equation}
where $\hat{\bm{\Lambda}}_{\eta=\pm}$ is given in Eq.~\eqref{eq.full.Lamb}.
\par
\subsection{Quantum kinetic equation for voltage-driven superconducotor}
\par
To evaluate the time evolution of the superconducting order parameter $\Delta_j(t>0)$ after the time-dependent voltage $V_0+\Delta V(t)$ is applied to the system, we derive the equation of motion of the equal-time lesser Green's function $\hat{\bm{G}}^<(t) \equiv \hat{\bm{G}}^<(t, t)$. [Note that $\hat{\bm{G}}^<(t)$ is directly related to the superconducting order parameter $\Delta_j(t)$ via Eq.~\eqref{eq.OP}.] Substituting the self-energy corrections in Eqs.~\eqref{eq.selfR.int}, \eqref{eq.selfL.int}, \eqref{eq.selfR.env}, and \eqref{eq.selfL.env} to the Dyson equations \eqref{eq.Dyson.GR} and \eqref{eq.Dyson.GL}, we obtain (For the derivation, see Appendix~\ref{eq.sec.app.QKE}.)
\begin{align}
&
i\partial_t \hat{\bm{G}}^<(t) =
\big[\hat{\bm{\m{H}}}_{\rm BdG}(t), \hat{\bm{G}}^<(t)\big]_-\notag\\
&\hspace{3cm}
 -4i\gamma \hat{\bm{G}}^<(t) 
-\hat{\bm{\Pi}}(t) -\hat{\bm{\Pi}}^\dagger(t),
\label{eq.main}
\end{align}
where
\begin{align}
&\hat{\bm{\m{H}}}_{\rm BdG}(t) = \hat{\bm{\m{H}}}_0 -\hat{\bm{\Delta}}(t)
\label{eq.HBdG},\\[4pt]
&\hat{\bm{\Pi}}(t)= \int_{-\infty}^{\infty} dt_1\hat{\bm{G}}^{\rm r}(t, t_1) \hat{\bm{\Sigma}}^<_{\rm lead} (t_1,t),
\label{eq.Pi}
\end{align}
and $\hat{\bm{\Pi}}(t)$ in Eq.~\eqref{eq.Pi} involves the retarded Green's function $\hat{\bm{G}}^{\rm r}(t, t')$, so that the quantum kinetic equation~\eqref{eq.main} is solved together with the Dyson equation~\eqref{eq.Dyson.GR}.
\par
We note that the first term on the right-hand side in Eq.~\eqref{eq.main} represents the unitary time evolution. When we only retain this term by setting $\gamma=0$, which physically means cutting off the couplings between the superconductor and the normal-metal leads, Eq.~\eqref{eq.main} is reduced to
\begin{equation}
i\partial_t \hat{\bm{G}}^<(t) =
\big[\hat{\bm{\m{H}}}_{\rm BdG}(t), \hat{\bm{G}}^<(t)\big]_-.
\label{eq.TDBdG}
\end{equation}
This equation is equivalent to the so-called time-dependent Bogoliubov–de Gennes (TDBdG) equation \cite{KettersonBook, Andreev1823}, which is widely used in studying the dynamics of {\it isolated} superconductors. In this sense, Eq.~\eqref{eq.main} is an extension of the TDBdG equation to an {\it open} superconductor in the N-S-N junction. 
\par
We also note that  Eq.~\eqref{eq.main} is an integro-differential equation that depends on the past information through $\hat{\bm{\Pi}}(t)$ in Eq.~\eqref{eq.Pi}. This so-called memory effect comes from the couplings with the non-Markovian reservoirs \cite{Collado2019, Collado2020}: The $\omega$ dependence of the energy distribution function $f (\omega)$ in the $\alpha$ reservoirs makes the lesser self-energy $\hat{\bm{\Sigma}}^<_{\rm lead}(t, t')$ in Eq.~\eqref{eq.selfL.env} nonlocal in time, which results in the non-Markovian term $\hat{\bm{\Pi}}(t)$. 
\par
\par
\subsection{Auxiliary-mode expansion}
\par
We numerically compute the time evolution of the superconducting order parameter $\Delta_j(t)$ under the voltage $V(t)=V_0 +\Delta V(t)\Theta(t)$, by solving Eq.~\eqref{eq.main} together with the Dyson equation~\eqref{eq.Dyson.GR}. Although similar coupled equations have already been solved in Refs.~\cite{Collado2019, Collado2020} by carrying out the integral of $\Pi(t)$ in Eq.~\eqref{eq.Pi} at each time step, these previous papers deal only with a much simpler spatially homogeneous superconducting case at $T_{\rm env}=0$ and $V(t)=0$. However, in the present case, the non-Markovian nature of Eq.~\eqref{eq.main} makes such a direct computation very challenging.
\par
To circumvent the problem, we extend the auxiliary-mode expansion technique, developed in the study of the time-dependent electron transport in a normal-metal device \cite{Croy2009, Croy2011, Croy2012, Popescu2018, Lehmann2018, Tuovinen2023}, to the present superconducting junction. This technique allows us to convert the integro-differential equation~\eqref{eq.main} into a system of ordinary differential equations that are suitable for numerical calculations.
\par
The main idea of the technique is performing the $\omega$ integral in Eq.~\eqref{eq.selfL.env} by using the residue theorem. To this end, we expand the Fermi-Dirac function $f(\omega -\eta e V_0/2)$ as
\begin{equation}
f\left(\omega -\eta \frac{eV_{0}}{2}\right) \simeq 
\frac{1}{2} -T_{\rm env} \sum_{n=1}^{N_{\rm F}}\left[
\frac{r_n}{\omega -\chi_{\eta, n}} +
\frac{r_n}{\omega -\chi^*_{\eta, n}} \right],
\label{eq.Fermi.ex}
\end{equation}
where the summation is taken over simple poles of
\begin{equation}
\chi_{\eta, n} = \eta \frac{eV_{0}}{2} + i \chi_{n},
\end{equation}
with $r_n$ being their residues. The choice of ($\chi_n, r_n$) is not unique. The well-known example is the Matsubara expansion \cite{MahanBook} with the Matsubara frequency $\chi_n = [2n+1]\pi T_{\rm env}$ and $r_n=1$. Instead of using this expansion, we use the Pad\'{e} expansion \cite{Ozeki2007, Karrasch2010}, which converges much faster than the Matsubara expansion. (We checked that $N_{\rm F}=10$ is sufficient in the case of the Pad\'{e} expansion.) For details of this expansion, see Appendix~\ref{app.pole}.
\par
Substituting the expanded Fermi-Dirac function in Eq.~\eqref{eq.Fermi.ex} into Eq.~\eqref{eq.selfL.env} and using the residue theorem, we have
\begin{equation}
\hat{\bm{\Sigma}}^<_{\rm lead}(t, t') =\sum_{\eta=\pm}
\left[i\gamma \delta(t-t') +\sum_{n=1}^{N_{\rm F}} \phi_{\eta, n}(t,t')\right]\hat{\bm{1}}, 
\label{eq.exp.selfL}
\end{equation}
with
\begin{align}
\phi_{\eta, n}(t,t') 
&=
2\gamma T_{\rm env} r_n e^{-i \chi_{\eta, n}(t-t')}
\notag\\
&\hspace{1.5cm}\times
\exp\left(-i \eta \int_{t'}^t dt_1 \frac{e\Delta V(t_1)}{2} \right).
\end{align}
We then substitute the expanded self-energy in Eq.~\eqref{eq.exp.selfL} into $\hat{\bm{\Pi}}(t)$ in Eq.~\eqref{eq.Pi}, which reads
\begin{equation}
\hat{\bm{\Pi}}(t)= \gamma\hat{\bm{1}} +
\sum_{\eta=\pm} \sum_{n=1}^{N_{\rm F}}\hat{\bm{\Pi}}_{\eta, n}(t). 
\label{eq.exp.Pi}
\end{equation}
Here, we use $\hat{\bm{G}}^{\rm r}(t, t) =-i \hat{\bm{1}}/2$ and introduce
\begin{equation}
\hat{\bm{\Pi}}_{\eta, n}(t)=
\int_{-\infty}^\infty dt_1 \hat{\bm{G}}^{\rm r}(t,t_1) \phi_{\eta, n}(t_1, t),
\label{eq.Pi.eta.n}
\end{equation}
whose equation of motion is found 
\begin{align}
&
i\partial_t \hat{\bm{\Pi}}_{\eta, n}(t)= 
2\gamma T_{\rm env} r_n \hat{\bm{1}} 
\notag\\[4pt]
&\hspace{1cm}+
\left[\hat{\bm{\m{H}}}_{\rm BdG}(t) -2i\gamma -\chi_{\eta, n} -\eta \frac{e\Delta V(t)}{2}\right]
\hat{\bm{\Pi}}_{\eta, n}(t).
\label{eq.EOM.exp.pi}
\end{align}
In deriving Eq.~\eqref{eq.EOM.exp.pi}, we have used Eq.~\eqref{eq.EOM.Gr}, which is equivalent to the Dyson equation~\eqref{eq.Dyson.GR}, and 
\begin{equation}
i\partial_{t'} \phi_{\eta, n}(t, t') = -
\left[\chi_{\eta, n} +\eta \frac{e\Delta V(t')}{2} \right] \phi_{\eta, n} (t, t').
\end{equation}
Substituting Eq.~\eqref{eq.exp.Pi} into Eq.~\eqref{eq.main}, we arrive at
\begin{align}
&
i\partial_t \hat{\bm{G}}^<(t) =
\big[\hat{\bm{\m{H}}}_{\rm BdG}(t), \hat{\bm{G}}^<(t)\big]_- 
\notag\\
&\hspace{0.5cm}
-4i\gamma\hat{\bm{G}}^<(t) -2\gamma \hat{\bm{1}} -\sum_{\eta=\pm} \sum_{n=1}^{N_{\rm F}}\left[\hat{\bm{\Pi}}_{\eta, n}(t)+\hat{\bm{\Pi}}^\dagger_{\eta, n}(t)\right].
\label{eq.main2}
\end{align}
Thus, the original coupled Dyson equation~\eqref{eq.Dyson.GR} with the (computationally challenging) integro-differential equation~\eqref{eq.main} is now replaced by the set of the ordinary differential equations given in Eqs.~\eqref{eq.EOM.exp.pi} and \eqref{eq.main2}.
\par
We numerically solve Eqs.~\eqref{eq.EOM.exp.pi} and \eqref{eq.main2}, by using the fourth-order Runge-Kutta method with sufficiently small time steps. At each time step, $\Delta_j(t)$ in $\hat{\bm{\m{H}}}_{\rm BdG}(t)$ is evaluated from $\hat{\bm{G}}^<(t)$ via Eq.~\eqref{eq.OP}, to proceed to the next time step. The initial condition at $t=0$ is given by (see Sec.~\ref{sec.NESS})
\begin{widetext}
\begin{align}
\hat{\bm{G}}^<(t=0) 
&=\int_{-\infty}^\infty \frac{d\omega}{2\pi}  \hat{\bm{G}}^<_{\rm NESS}(\omega)
= \hat{\bm{W}} \hat{\bm{\Lambda}} \hat{\bm{W}}^\dagger
,\\[4pt]
\hat{\bm{\Pi}}^<_{\eta, n}(t=0) 
&=\int_{-\infty}^\infty \frac{d\omega}{2\pi} \hat{\bm{G}}^{\rm r}_{\rm NESS}(\omega)\phi_{\eta, n}(\omega)
=2\gamma T_{\rm env} r_n \bm{W}
\begin{pmatrix}
[\chi_{\eta, n}+2i\gamma -E_1]^{-1} &  \\
 & \ddots \\
&& [\chi_{\eta, n}+2i\gamma -E_{2N}]^{-1} 
\end{pmatrix}
\bm{W}^\dagger.
\end{align}
Here, $\hat{\bm{W}}$ and $E_{j=1,\cdots, 2N}$ are introduced in Eq.~\eqref{eq.def.Bogo}, and $\hat{\bm{\Lambda}}$ is defined in Eq.~\eqref{eq.def.Lamb}.
\end{widetext}
\par
\par
\subsection{Momentum-space formalism}
\par
At the end of this section, we map the real-space formalism explained in Sec.~\ref{sec.NESS} onto the momentum-space formalism for later convenience.
\par
In the case of the spatially uniform superconducting state ($\Delta_{j=1,\cdots,N}=\Delta$), it is convenient to Fourier-transform the field operator as
\begin{equation}
c_{\bm{k}, \sigma} = \sum_{\bm{r}_j} e^{-i\bm{r}_j \cdot \bm{k}} c_{\bm{r}_j, \sigma},
\end{equation}
where $\bm{r}_j=(x_j, y_j)$ is the position vector of $j$th lattice site. As shown in Appendix.~\ref{sec.app.gapeq}, by evaluating the lesser Green's function in momentum space, we obtain the nonequilibrium BCS gap equation given by \cite{Kawamura2022}
\begin{equation}
\scalebox{0.98}{$\displaystyle
\frac{1}{U} = \sum_{\bm{k}} \int_{-\infty}^\infty \frac{d\omega}{2\pi} \frac{4\gamma\omega\big[1 -f(\omega-eV_0/2)-f(\omega+eV_0/2)\big]}{\big[(\omega -E_{\bm{k}})^2 +4\gamma^2\big] \big[(\omega +E_{\bm{k}})^2 +4\gamma^2\big]}. $}
\label{eq.noneq.gap}
\end{equation}
Here, 
\begin{equation}
E_{\bm{k}} = \sqrt{\ep^2_{\bm{k}} +\Delta^2}
\label{eq.Bogo.Ep}
\end{equation}
is the Bogoliubov excitation energy with
\begin{equation}
\ep_{\bm{k}} = -2t_\parallel \big[ \cos(k_x) +\cos(k_y) \big] -\mu_{\rm sys},
\end{equation}
being the kinetic energy of an electron (where the lattice constant is taken to be unity, for simplicity). We solve the gap equation~\eqref{eq.noneq.gap} self-consistently to obtain $\Delta$ for a given parameter set ($\gamma, V_0, T_{\rm env}$). 
\par
Particularly in the thermal-equilibrium limit ($\gamma\to +0$ and $V_0=0$), Eq.~\eqref{eq.noneq.gap} is reduced to
\begin{equation}
\frac{1}{U} = \sum_{\bm{k}} \frac{1}{2E_{\bm{k}}}\tanh\left(\frac{E_{\bm{k}}}{2T_{\rm env}}\right),
\end{equation}
which is just the same form as the ordinary BCS gap equation \cite{KettersonBook, SchriefferBook}, when one interprets $T_{\rm env}$ as the temperature in the system. In this sense, Eq.~\eqref{eq.noneq.gap} may be interpreted as a nonequilibrium extension of the BCS gap equation.
\par
We can also work in momentum space, when the system is in the normal state ($\Delta_{j}=0$). As shown by Kadanoff and Martin (KM) \cite{Kadanoff1961, KadanoffBook}, when the particle-particle scattering $T$-matrix $\chi^{\rm r}(\bm{q}, \nu)$ in the normal phase has a pole at $(\bm{q}, \nu)=(\bm{Q}, 0)$, the normal state becomes unstable (Cooper instability) and the superconducting transition occurs. By extending the KM theory to the nonequilibrium case, we obtain the $T_{\rm env}^{\rm c}$-equation that determines the boundary between the normal phase and the superconducting phase.
\par
Evaluating the nonequilibrium $T$-matrix $\chi^{\rm r}(\bm{q}, \nu)$ by using the nonequilibrium Green's function technique, we obtain the $T_{\rm env}^{\rm c}$-equation from the KM condition $[\chi^{\rm r}(\bm{q}=\bm{Q}, \nu=0)]^{-1} =0$ as \cite{Kawamura2020JLTP, Kawamura2020, Kawamura2023}
\begin{align}
&
\frac{1}{U} =\sum_{\bm{k}} \int_{-\infty}^\infty \frac{d\omega}{2\pi} 
\notag\\
&\hspace{0.3cm}\times
\frac{4\gamma \big[\omega -\ep^{\rm a}_{\bm{k}, \bm{Q}}\big]\big[1 -f(\omega-eV_0/2) -f(\omega+eV_0/2)\big]}{\big[(\omega -\ep_{\bm{k}+\bm{Q}/2})^2 +4\gamma^2\big]\big[(\omega+\ep_{-\bm{k}+\bm{Q}/2})^2 +4\gamma^2\big]}, 
\label{eq.Thouless}
\end{align}
where
\begin{equation}
\ep^{\rm a}_{\bm{k}, \bm{Q}} = \frac{1}{2}\big[\ep_{\bm{k}+\bm{Q}/2} -\ep_{-\bm{k}+\bm{Q}/2}\big].
\end{equation}
We summarize the derivation of Eq.~\eqref{eq.Thouless} in Appendix.~\ref{sec.app.KM}. In Eq.~\eqref{eq.Thouless}, the momentum $\bm{Q}$ is determined so as to obtain the highest $T_{\rm env}^{\rm c}$. When $\bm{Q}=0$, the BCS-type uniform superconducting state is realized. Indeed, the $T_{\rm env}^{\rm c}$-equation~\eqref{eq.Thouless} with $\bm{Q}=0$ just equals the gap equation~\eqref{eq.noneq.gap} with $\Delta=0$. On the other hand, the solution with $\bm{Q}\neq 0$ describes an inhomogeneous superconducting state, being characterized by a spatially oscillating order parameter (symbolically written as $\Delta_{\bm{r}_j} = \Delta e^{i\bm{Q}\cdot \bm{r}_j}$). This is analogous to the Fulde–Ferrell–Larkin–Ovchinnikov (FFLO) state in a superconductor under an external magnetic field \cite{Fulde1964, Larkin1964, Matsuda2007}.
\par
\section{Nonequilibrium phase diagram of the voltage-driven superconductor \label{sec.result}}
\par
\begin{figure}[t]
\centering
\includegraphics[width=8.2cm]{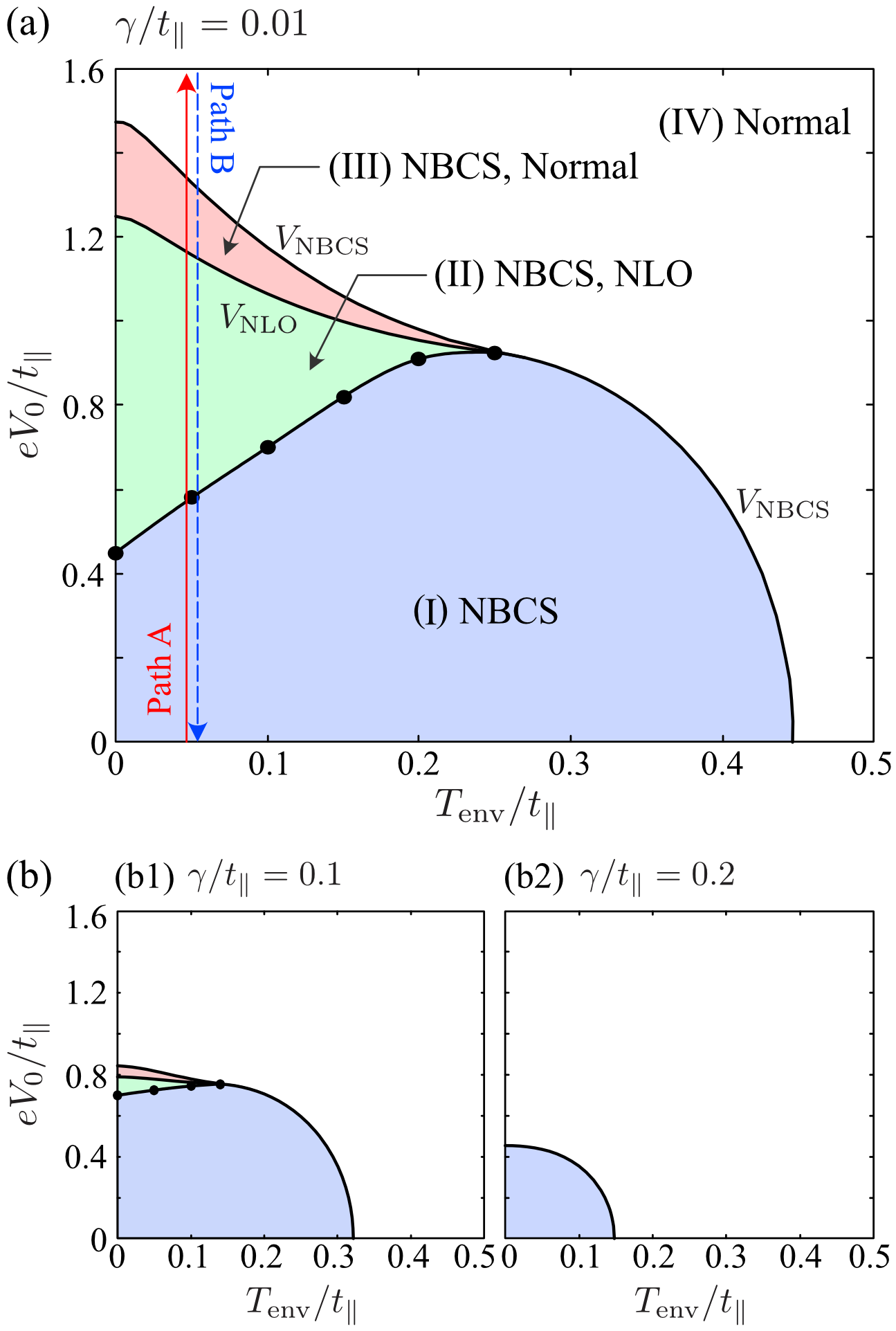}
\caption{(a) Nonequilibrium phase diagram of the voltage-driven superconductor, when $\gamma/t_\parallel =0.01$. NBCS and NLO are the nonequilibrium BCS and Larkin-Ovchinnikov state, respectively. While NBCS is a uniform superconducting state, NLO is a nonuniform superconducting state with a spatially oscillating order parameter. The system exhibits bistability in regions II and III. When the system enters these regions from region I, NBCS is realized in both the regions. On the other hand, the system relaxes to NLO (normal state) in region II (III), when entering from region IV. $V_0=V_{\rm NBCS}$ (boundaries between region I and region IV, as well as region III and region IV) and $V_0=V_{\rm NLO}$ (boundary between region II and region III) are, respectively, obtained by solving Eqs.~\eqref{eq.noneq.gap} and \eqref{eq.Thouless} (see the main text for details). (b) Effects of the system-lead coupling strength $\gamma$ on the phase diagram.}
\label{Fig.phase.diagram}
\end{figure}
\par
We now explore the nonequilibrium phase diagram of the voltage-driven superconductor. Hereafter, we use the hopping amplitude $t_\parallel$ as the energy unit. We also set $\mu_{\rm sys}=0$ and $U/t_\parallel =3$. We focus on the spatial dependence of the order parameter $\Delta_j$ along the $x$ axis in a two-dimensional square lattice with $L_x\times L_y = 101\times 11$ sites.
\par
The main findings are as follows: 
\begin{enumerate}
\item When a constant bias voltage is applied, the system always relaxes to a certain steady state; that is, the system never realizes a time-periodic state as seen in a voltage-driven superconducting wire \cite{Langer1967, Skocpol1974, Kramer1977, Ivlev1984, Bezryadin2000, Vodolazov2003, Yerin2013, Yerin2013_2}. The phase diagram in Fig.~\ref{Fig.phase.diagram} summarizes the steady states of the system.
\item In region II of the phase diagram, a nonuniform superconducting state with a spatially oscillating order parameter, which we call the nonequilibrium Larkin-Ovchinnikov (NLO) state, is realized. The emergence of NLO is attributed to the nonequilibrium energy distribution function $f_{\rm neq}(\omega)$ in Eq.~\eqref{eq.fneq.**}.
\item In regions II and III of the phase diagram, the system exhibits bistability.  When the system enters these regions from region I, a uniform superconducting state, which we call the nonequilibrium BCS (NBCS) state, is realized in both regions. On the other hand, when the system enters these regions from region IV, NLO and the normal state are, respectively, realized in region II and region III.
\end{enumerate}
Below we discuss these findings in detail.
\par
\begin{figure}[t]
\centering
\includegraphics[width=8.6cm]{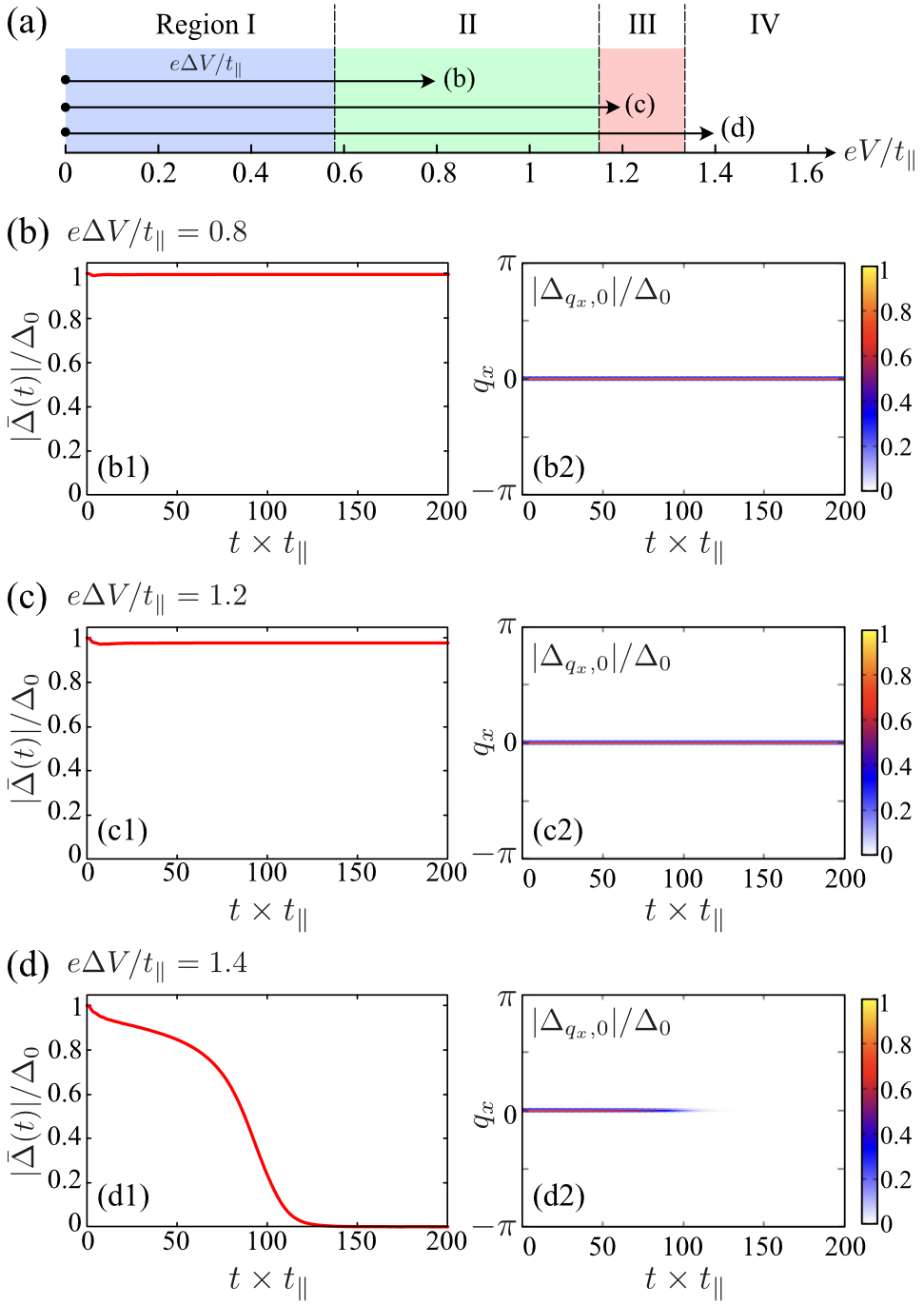}
\caption{Time evolution of the order parameter under the voltage $V(t)=\Delta V\Theta(t)$. The initial state at $t=0$ is the thermal equilibrium BCS state ($V_0=0$). We set $\gamma/t_\parallel =0.01$ and $T_{\rm env}/t_\parallel =0.05$. (a) Schematic diagram of how we quench the voltage. The regions I-IV are shown in Fig.~\ref{Fig.phase.diagram} (a). Panels (b)-(d) show the time evolution of $|\bar{\Delta}(t)|$ in Eq.~\eqref{eq.ave.Delta} and $\Delta_{q_x, q_y=0}(t)$ in Eq.~\eqref{eq.qx.Delta} for each voltage quench depicted in panel (a). Here, $\Delta_0$ is the order parameter in the case that the superconductor is isolated from the normal-metal leads and is in the BCS ground state.
}
\label{Fig.quench.from000}
\end{figure}
\par
\begin{figure}[t]
\centering
\includegraphics[width=8.2cm]{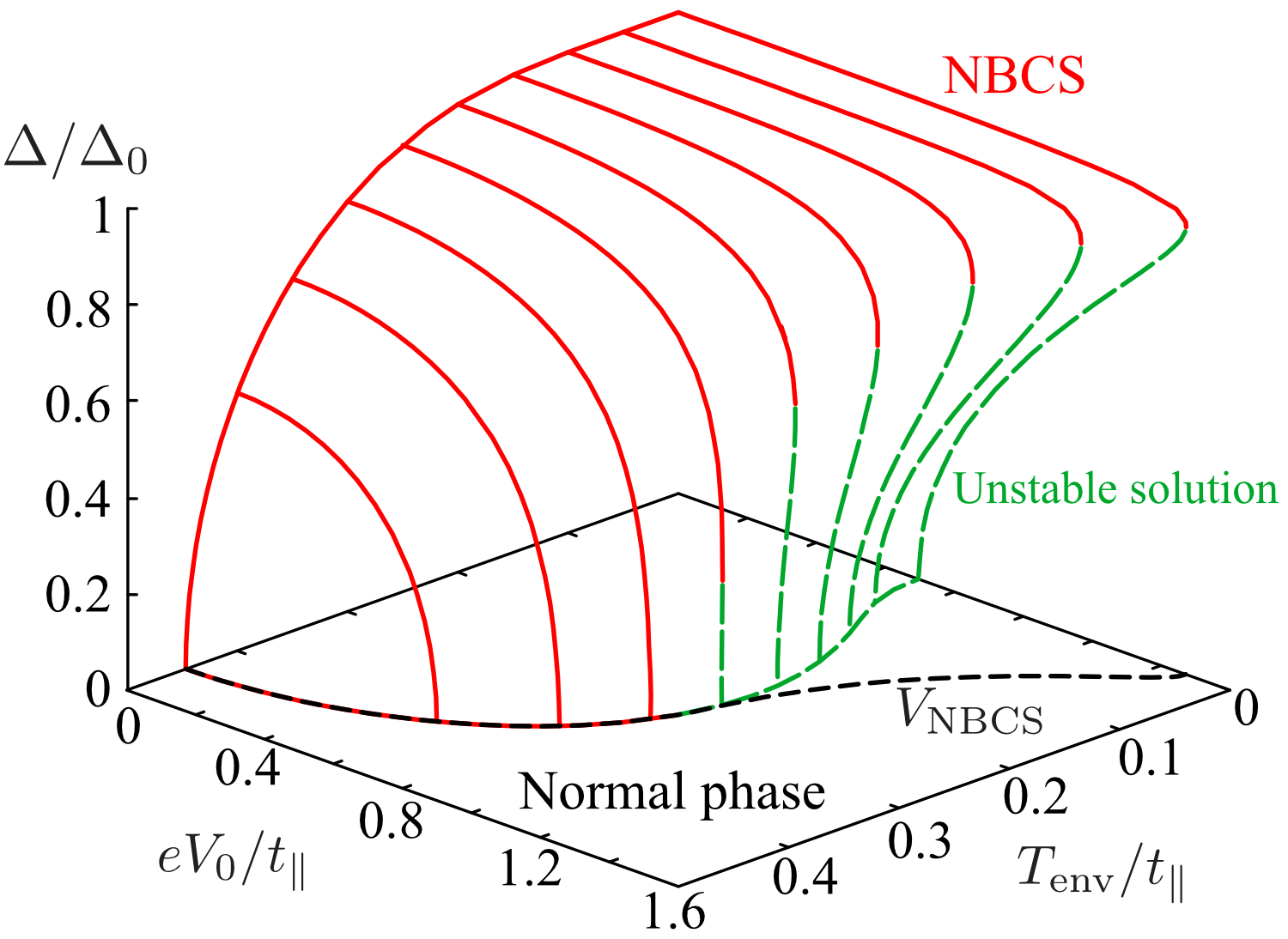}
\caption{Self-consistent solutions of the nonequilibrium gap equation~\eqref{eq.noneq.gap}, when $\gamma/t_\parallel =0.01$. $V_0=V_{\rm NBCS}$ gives the boundaries between region I and region IV, as well as region III and region IV in the nonequilibrium phase diagram in Fig.~\ref{Fig.phase.diagram}. While the solid line corresponds to NBCS, the dashed line corresponds to an unstable gapless superconducting state.
}
\label{Fig.Delta} 
\end{figure}
\par
First, we consider the case when the system is initially in the thermal equilibrium BCS state ($V_0= 0$) at $t=0$ and then driven out of equilibrium by the voltage $V(t)=\Delta V \Theta(t)$. Figure~\ref{Fig.quench.from000} summarizes the $\Delta V$ dependence of the time evolution of the order parameter obtained by solving the coupled differential equations~\eqref{eq.EOM.exp.pi} and \eqref{eq.main2}. In this figure, $|\bar{\Delta}(t)|$ and $\Delta_{q_x, q_y}$ are defined by
\begin{align}
& 	|\bar{\Delta}(t)| = \frac{1}{N} \sum_{j=1}^N |\Delta_j(t)|
\label{eq.ave.Delta}
,\\
& \Delta_{q_x, q_y}(t)= \frac{1}{L_x L_y} \sum_{j=1}^N \Delta_{j}(t) e^{i \bm{r}_j \cdot \bm{q}}.
\label{eq.qx.Delta}
\end{align}
Here, $|\bar{\Delta}(t)|$ is the spatial average of the order-parameter amplitude and $\Delta_{q_x, q_y}(t)$ is the order parameter in momentum $\bm{q}=(q_x, q_y)$ space. 
\par
Figure~\ref{Fig.quench.from000} shows that, when the voltage is quenched from $V_0=0$, the system always relaxes to a steady state. When the system enters regions II and III ($e\Delta V/t_\parallel = 0.8$ and $1.2$), the system relaxes to a uniform superconducting state (NBCS), where $|\Delta_{q_x, q_y}|$ has a peak only at $\bm{q}=0$, as shown in Figs.~\ref{Fig.quench.from000}(b2) and (c2). On the other hand,  Fig.~\ref{Fig.quench.from000}(d) shows that the superconducting state transitions to the normal state ($|\bar{\Delta}|=0$) when entering region IV ($e\Delta V /t_\parallel= 1.4$). To conclude, when the voltage is increased from $V_0=0$, NBCS is maintained in regions I-III and the system transitions to the normal state when entering region IV. 
\par
The boundary between region III and region IV can be easily obtained from the nonequilibrium gap equation~\eqref{eq.noneq.gap} \cite{Kawamura2022}. Figure~\ref{Fig.Delta} shows the temperature $T_{\rm env}$ and the voltage $V_0$ dependence of the solution $\Delta$ of the nonequilibrium gap equation~\eqref{eq.noneq.gap}, when $\gamma/t_\parallel =0.01$. As seen from this figure, superconducting solutions ($\Delta\neq 0$) vanish in the case when $V_0 >V_{\rm NBCS}$. Thus, $V_{\rm NBCS}$ is the critical voltage for the uniform superconducting state (NBCS), and $V_0=V_{\rm NBCS}$ gives the boundaries between region I and region IV, as well as region III and region IV, in the nonequilibrium phase diagram in Fig.~\ref{Fig.phase.diagram}.
\par
We note that the nonequilibrium gap equation~\eqref{eq.noneq.gap} has two solutions in the low-temperature regime $T_{\rm env}/t_\parallel\lesssim 0.35$, as shown in Fig.~\ref{Fig.Delta} \cite{Kawamura2022}. While the solid line corresponds to NBCS, the dashed line corresponds to a spatially uniform gapless superconducting state analogous to the  Sarma(-Liu-Wilczek) state discussed under an external magnetic field \cite{Sarma1963, Liu2003}. Since this gapless superconducting state is an unstable steady state \cite{Kawamura2022}, the system actually never relaxes to this state.
\par
\begin{figure}[t]
\centering
\includegraphics[width=8.6cm]{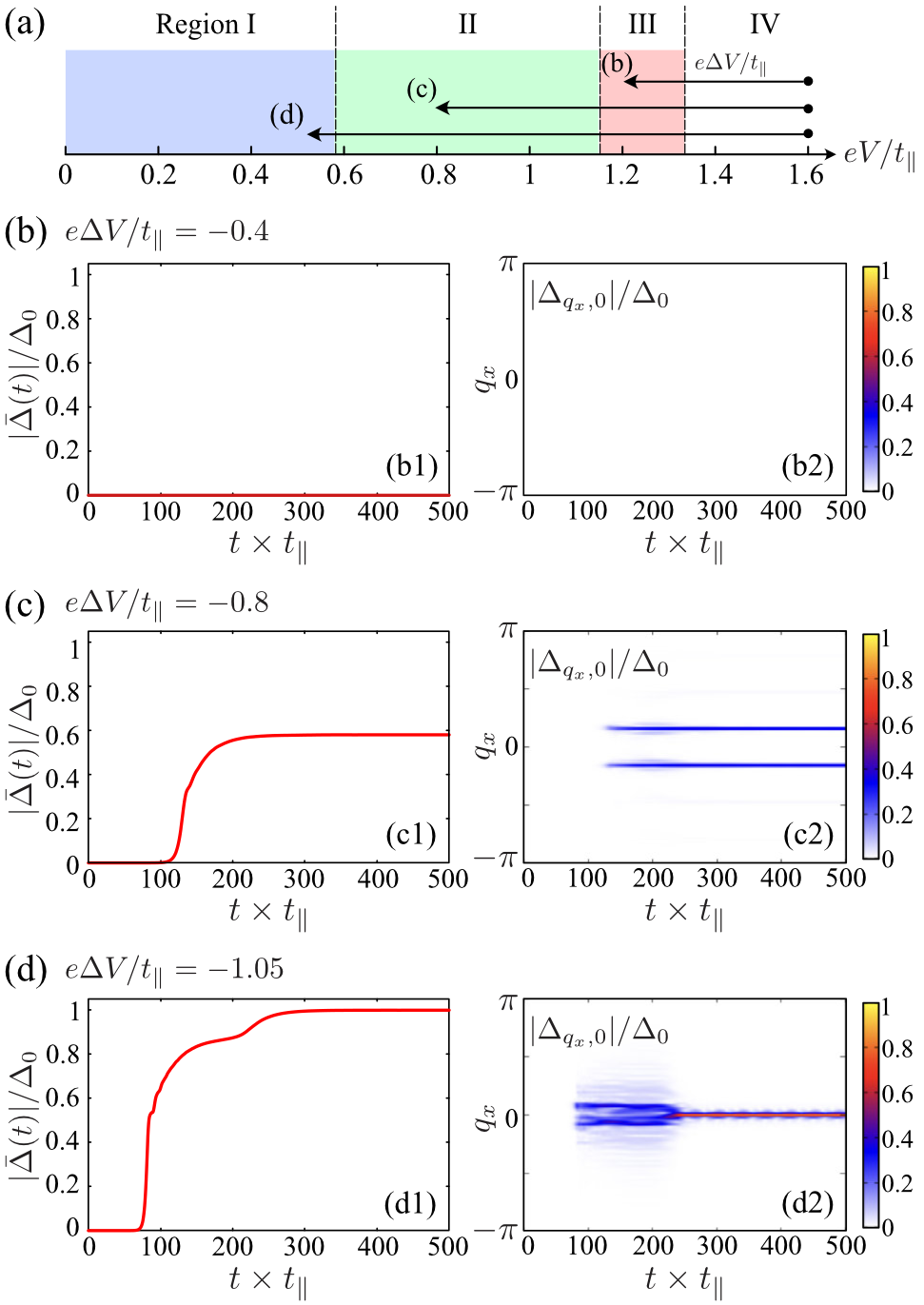}
\caption{Same plots as Fig.~\ref{Fig.quench.from000} for the voltage $e V(t)/t_\parallel=1.6 +e \Delta V\Theta (t)/t_\parallel$. The initial state at $t=0$ is the normal steady state ($eV_0/t_\parallel=1.6$).} 
\label{Fig.quench.from160}
\end{figure}
\par
Since we find that the system is in the normal steady state in region IV, we next discuss the time evolution of the order parameter $\Delta_j(t)$, when the voltage is decreases from region IV. (As a typical example, we set $eV_0/t_\parallel=1.6$.) To trigger the superconducting phase transition, we give a spatially random small amplitude $|\Delta_j(t=0)|/\Delta_0=10^{-5} \zeta_j$, as well as random phase $\theta_j(t=0)=2\pi \zeta_j$, (where $\zeta_j$ is a random number between $0$ and $1$) for the initial order parameter $\Delta_j(0)=|\Delta_j(0)|e^{i\theta_j(0)}$.
\par
As shown in Fig.~\ref{Fig.quench.from160}(b1), the order-parameter amplitude $|\bar{\Delta}|$ does not grow over time, when we decrease the voltage to enter region III ($e\Delta V/t_\parallel=-0.4$), meaning that the system remains in the normal steady state in region III. On the other hand, when the system enters regions I and II ($e\Delta V/t_\parallel =-1.05$ and $-0.8$), $|\bar{\Delta}|$ grows and the system undergoes the superconducting transition, as shown in Figs.~\ref{Fig.quench.from160}(c1) and (d1). 
\par
\begin{figure}[t]
\centering
\includegraphics[width=7.8cm]{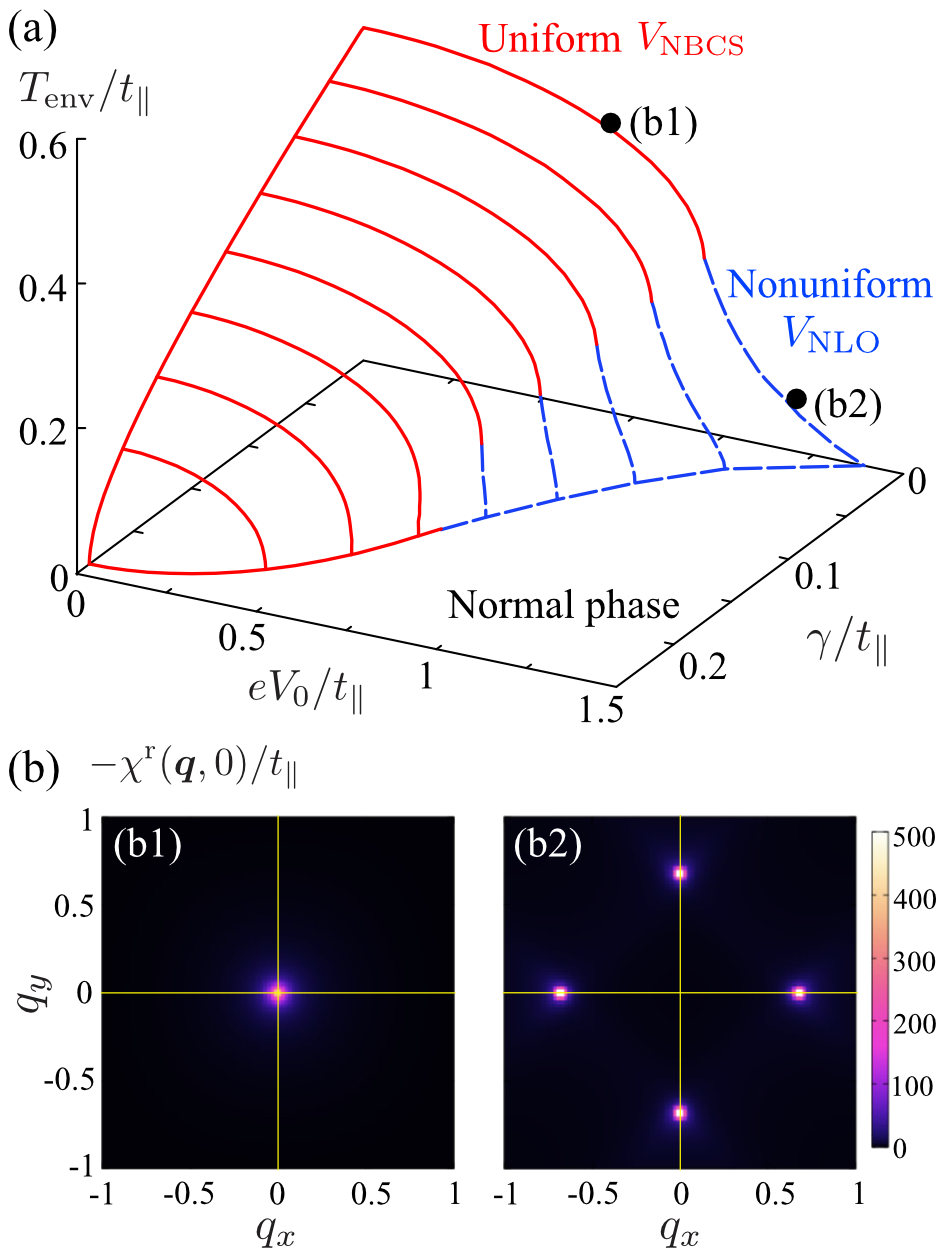}
\caption{(a) Superconducting transition line obtained from Eq.~\eqref{eq.Thouless}. The system transitions to a uniform NBCS (nonuniform NLO) on the solid $V_0=V_{\rm NBCS}$ (dashed $V_0=V_{\rm NLO}$) line. (b) The intensity of the $T$-matrix $\chi^{\rm r}(\bm{q}, 0)$ at the positions (b1) and (b2) shown in panel (a).}
\label{Fig.Thouless}
\end{figure}

The superconducting transition line (the boundary between region II and region III) shown in Fig.~\ref{Fig.Thouless}(a) is easily obtained from Eq.~\eqref{eq.Thouless} \cite{Kawamura2022}.  We see in Fig.~\ref{Fig.Thouless}(b) that the poles of the $T$-matrix $\chi^{\rm r}(\bm{q}, 0)$ appear at $\bm{q}=0$ on the solid line ($V_0=V_{\rm NBCS}$), whereas they appear at $\bm{q}=\bm{Q}(\neq 0)$ on the dashed line ($V_0=V_{\rm NLO}$). Thus, the solid  (dashed) line represents the phase transition from the normal state to uniform NBCS (nonuniform NLO). Hence, $V_0=V_{\rm NLO}$ gives the boundary between region II and region III in the nonequilibrium phase diagram in Fig.~\ref{Fig.phase.diagram}.
\par
\par
\begin{figure}[t]
\centering
\includegraphics[width=8.2cm]{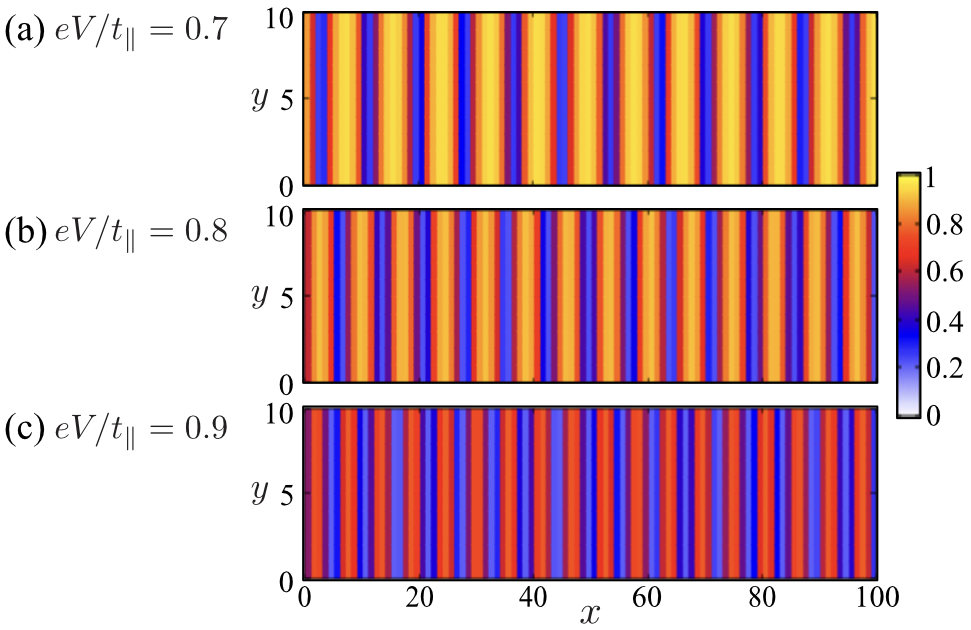}
\caption{The voltage dependence of the amplitude $|\Delta_j|$ of the NLO order parameter. The intensity is normalized by $\Delta_0$. We set $\gamma/t_\parallel =0.01$ and $T_{\rm env}/t_\parallel =0.05$.} 
\label{Fig.NLO.Q}
\end{figure}

Figure~\ref{Fig.quench.from160}(c2) shows that the system relaxes to the nonuniform superconducting state (NLO), where $|\Delta_{q_x, q_y=0}|$ has two peaks at $q_x = \pm Q$, in region II. The corresponding spatial profile of the order parameter, which can be symbolically written as $\Delta_{\bm{r}_j}=\Delta \cos(Qx_j)$, is shown in Fig.~\ref{Fig.NLO.Q}(b). The magnitude of $Q$ depends on the applied voltage. As shown in Fig.~\ref{Fig.NLO.Q}, the magnitude of $Q$ increases as the applied voltage increases. 
\par
\begin{figure}[t]
\centering
\includegraphics[width=8.4cm]{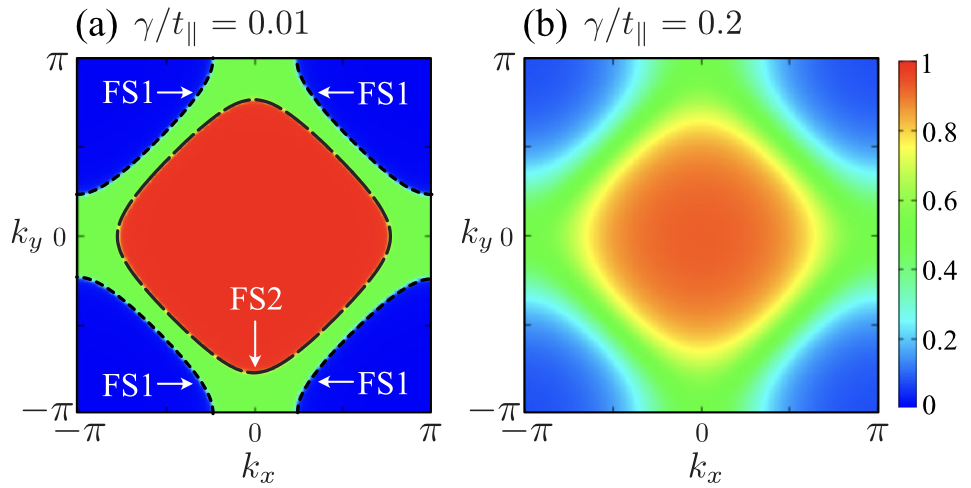}
\caption{Nonequilibrium momentum distribution $n_{\bm{k}}^{\rm neq}$ for different system-lead coupling strength $\gamma$. We set $T_{\rm env}=0$ and $eV_0/t_\parallel =1$ for both panels. Two Fermi edges (dashed and dotted lines) imprinted on $n_{\bm{k}}^{\rm neq}$ work like two ``Fermi surfaces" (FS1 and FS2) of different sizes.} 
\label{Fig.Fneq}
\end{figure}

The emergence of the LO-type superconducting state in region II can be attributed to the nonequilibrium energy distribution function $f_{\rm neq}(\omega)$ in Eq.~\eqref{eq.fneq.**} \cite{Kawamura2020JLTP, Kawamura2020, Kawamura2022, Kawamura2023}. Figure~\ref{Fig.Fneq}(a) shows the nonequilibrium momentum distribution $n_{\bm{k}}^{\rm neq}$ in Eq.~\eqref{eq.nk.noneq}. The two-step structure in $f_{\rm neq}(\omega)$ is taken over by the momentum distribution $n_{\bm{k}}^{\rm neq}$, creating the two Fermi edges, as shown in Fig.~\ref{Fig.Fneq}(a). These Fermi edges provide two effective ``Fermi surfaces" (FS1 and FS2) of different sizes, which induce the FFLO-type Cooper pairings with nonzero center-of-mass momentum. This mechanism is quite analogous to the ordinary thermal-equilibrium FFLO state induced by the Zeeman splitting between the spin-$\up$ and spin-$\down$ Fermi surfaces under an external magnetic field \cite{Fulde1964, Larkin1964}. However, in the nonequilibrium case, the spin-$\up$ and spin-$\down$ Fermi surfaces are exactly the same due to the absence of a Zeeman field. Instead, each spin component has two ``Fermi surfaces" of different sizes.
\par
\par
The voltage dependence of the NLO order parameter $\Delta_{\bm{r}_j}=\Delta\cos(Q x_j)$ shown in Fig.~\ref{Fig.NLO.Q} can be understood by using the effective ``Fermi surfaces": Since the magnitude of $Q$ physically implies a nonzero center-of-mass momentum of a Cooper pair, it depends on the difference in the size of FS1 and FS2. Because the position of FS1 and FS2 in momentum space is determined by the Fermi levels of left and right normal-metal leads, the size difference increases as the voltage increases. Thus,  the magnitude of $Q$ also increases as the voltage increases. 
\par
When the system enters region I from region IV ($e \Delta V/t_\parallel = -1.05$), although the NLO-like nonuniform superconducting state temporarily appears, the system eventually relaxes to the uniform superconducting state (NBCS), as shown in Fig.~\ref{Fig.quench.from160}(d2).
\par
Summarizing the results of Figs.~\ref{Fig.quench.from000} and \ref{Fig.quench.from160}, we arrive at the nonequilibrium phase diagram in Fig.~\ref{Fig.phase.diagram}(a). In regions I and IV, the system always relaxes to the uniform superconducting state (NBCS) and the normal state, respectively. On the other hand, the steady state realized in regions II and III depends on how we tune the voltage: As the voltage is increased from $V_0=0$, NBCS is maintained both in regions II and III, as shown in Figs.~\ref{Fig.quench.from000}(b) and (c). As the voltage is decreased from region IV, on the other hand, the system is in the normal state in region III and transitions to NLO in region II, as shown in Figs.~\ref{Fig.quench.from160}(b) and (c). We briefly note that once the system reaches the steady state, the phase $\theta_j(t)$ of the order parameter is independent of time, as is the order-parameter amplitude $|\bar{\Delta}(t)|$.
\par
As seen from Fig.~\ref{Fig.phase.diagram}(b), region II (where NLO is realized) is strongly suppressed as the system-lead coupling strength $\gamma$ increases. This is because the two Fermi edges imprinted on the nonequilibrium momentum distribution $n_{\bm{k}}^{\rm neq}$, which are the key factors inducing NLO, become obscure as $\gamma$ increases, as shown in Fig.~\ref{Fig.Fneq}(b). As $\gamma$ further increases, the NBCS phase (region I) in Fig.~\ref{Fig.phase.diagram}(b2) also disappears and the system does not show the superconducting transition. 
\par
\par
We note that the electron-electron, the electron-phonon, and the electron-impurity scattering would also make the two Fermi edges imprinted on $n_{\bm{k}}^{\rm neq}$ obscure and suppress NLO. Thus, to realize NLO, the applied voltage (which is typically about the superconducting gap $\Delta_0$) has to be large enough compared to the linewidth arising from these scattering processes, as well as the linewidth $\gamma$ due to the system-lead couplings. Since the linewidth due to the electron-electron and the electron-phonon scattering is much smaller than the superconducting gap in typical BCS superconductors \cite{Kaplan1976}, this condition is satisfied, when  $\gamma \ll \Delta_0$ and the superconductor is sufficiently clean (which is also known to be important for realizing the thermal-equilibrium FFLO state in a superconductor under an external magnetic field \cite{Matsuda2007}). 
\par
\begin{figure}[t]
\centering
\includegraphics[width=8.5cm]{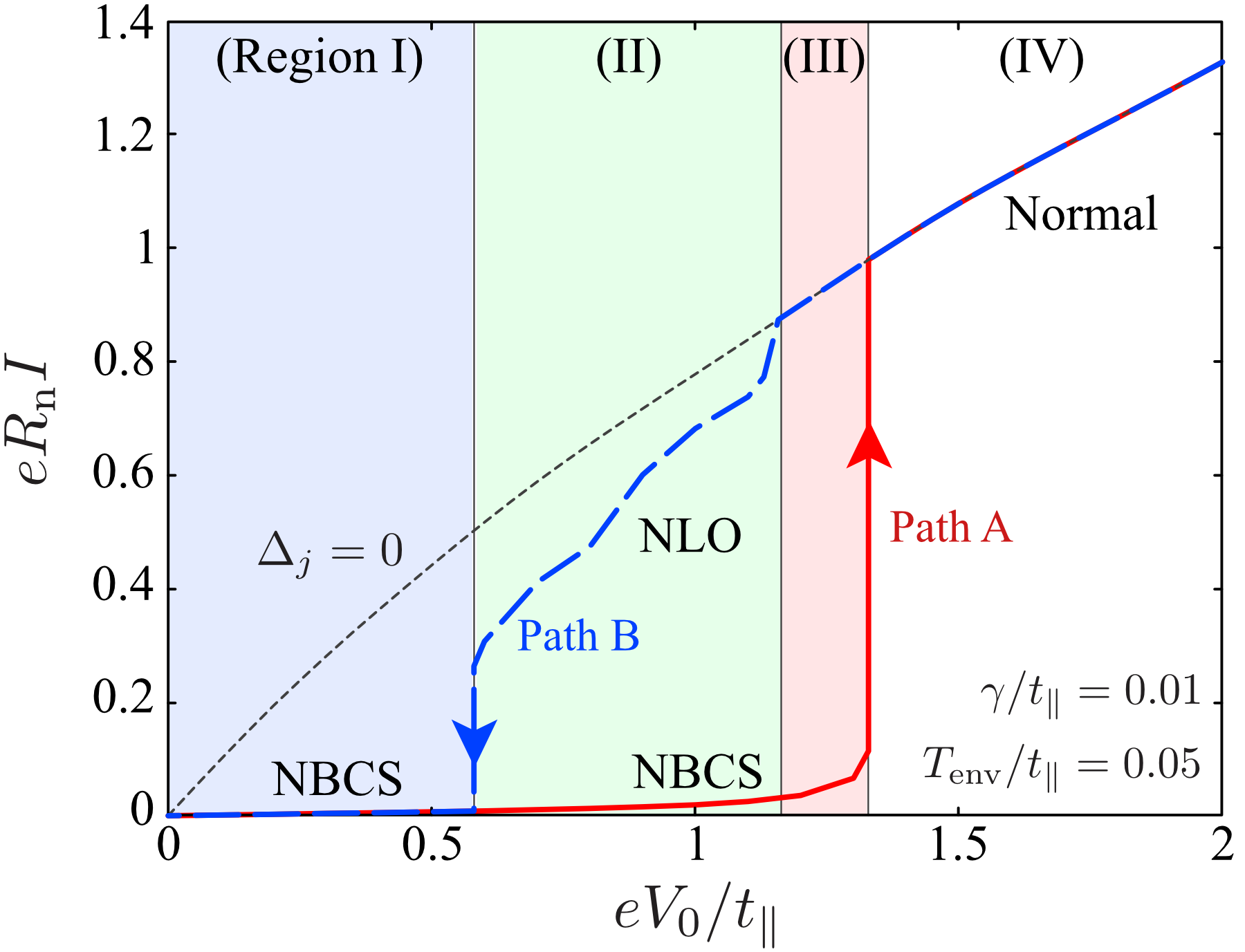}
\caption{Voltage-current characteristic of the N-S-N junction. The solid (dashed) line shows the steady-state current $I$ through the junction, when we adiabatically vary the voltage along path A (path B) in Fig.~\ref{Fig.phase.diagram}(a). The dotted line shows the result when the system is the normal state ($\Delta_j=0$). $R_{\rm n}$ is the normal resistance of the junction.
} 
\label{Fig.current}
\end{figure}

The bistability in the small-$\gamma$ regime leads to hysteresis in the voltage-current characteristic of the junction. Figure~\ref{Fig.current} shows the voltage dependence of the steady-state current $I$ in Eq.~\eqref{eq.sym.J}, when the voltage is adiabatically increased (decreased) along path A (path B) in Fig.~\ref{Fig.phase.diagram}(a). With increasing the voltage along path A, NBCS is maintained in regions II and III. In this case, the current $I$ is strongly suppressed due to the superconducting energy gap \cite{Demers1971, Griffin1971, Blonder1982} until the system enters region IV and transitions to the normal state. With decreasing the voltage along path B, the current $I$ is suppressed, when the system enters region II and transitions to NLO. However, since the NLO order parameter has spatial line nodes (which provide the paths for the single-electron tunneling), the current $I$ is larger than the NBCS case. Figure~\ref{Fig.current} indicates that a voltage-current measurement may be useful for the observation of the bistability, as well as NLO.
\par
\par
At the end of this section, we briefly discuss how the above results are affected by the boundary of the present model. So far, we focused only on the modulation of the order parameter along the $x$ axis by setting $L_x=101 > L_y =11$. Here we set $L_x=L_y$ to explore the possibility of a two-dimensional oscillation of the superconducting order parameter.
\par
\begin{figure}[t]
\centering
\includegraphics[width=8.2cm]{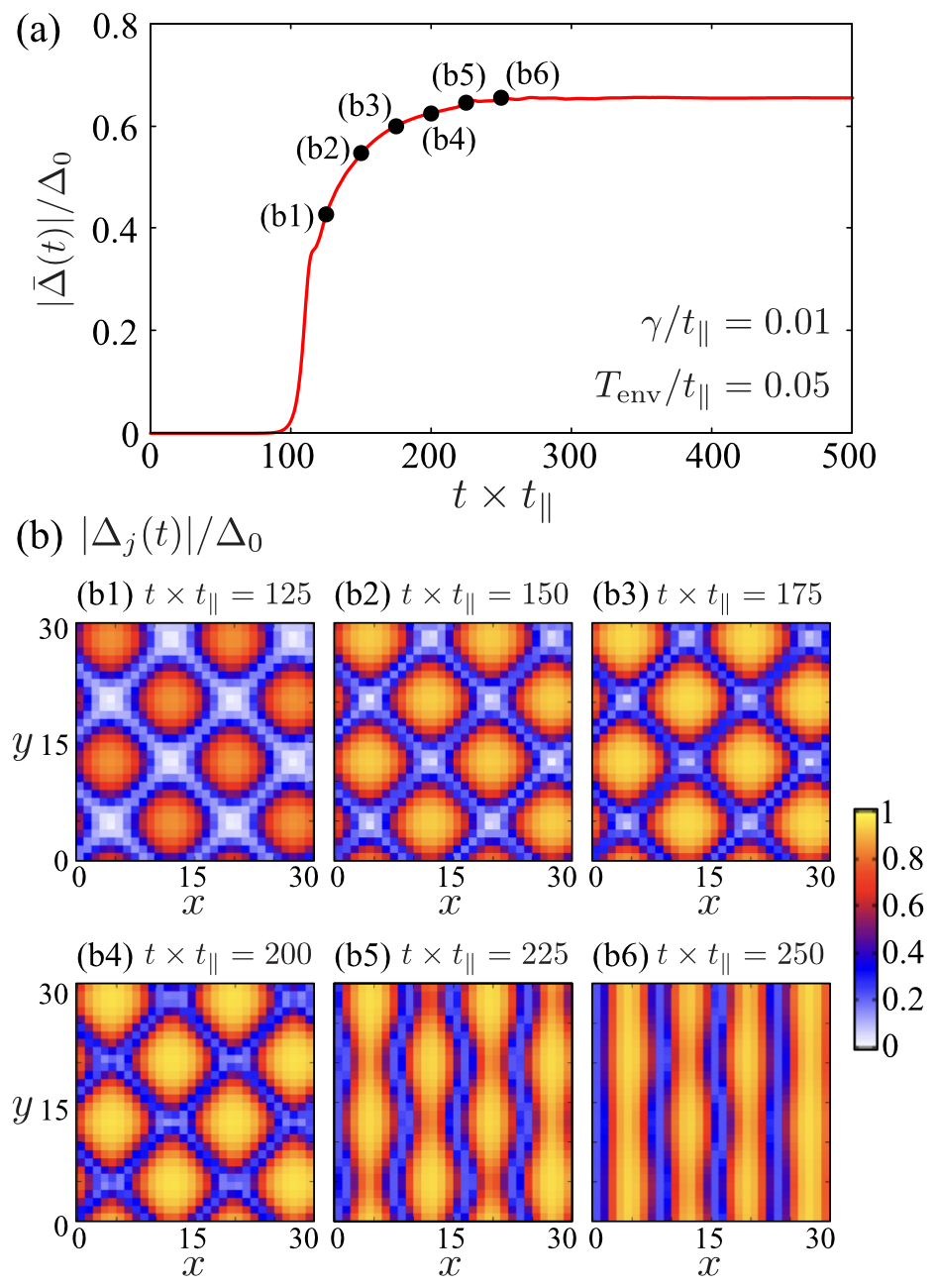}
\caption{Time evolution of the order parameter $\Delta_j(t)$ under the voltage $eV(t)/t_\parallel =1.6-0.8\Theta(t)$. We set $\gamma/t_\parallel =0.01$ and $T_{\rm env}/t_\parallel =0.05$, and $L_x \times L_y = 31\times 31$. At $t=0$, the system is in the normal steady state  [region IV in Fig.~\ref{Fig.phase.diagram}(a)]. (a) The time evolution of $|\bar{\Delta}(t)|$ in Eq.~\eqref{eq.ave.Delta}. (b) The spatial profile of the order-parameter amplitude $|\Delta_j(t)|$ at each point shown in panel (a).}
\label{Fig.NLO2}
\end{figure}
\par
Figure~\ref{Fig.NLO2} shows the time evolution of the order parameter under the voltage $eV(t)/t_\parallel=1.6 -0.8\Theta(t)$. The initial state at $t=0$ is the normal steady state [region IV in Fig.~\ref{Fig.phase.diagram}(a)]. Here we set $L_x\times L_y = 31\times 31$. As seen from Fig.~\ref{Fig.NLO2}(a), the order-parameter amplitude grows over time and the system transitions to a superconducting state. Figure~\ref{Fig.NLO2}(b6) shows that the system eventually relaxes to NLO, which is the same as the result obtained in the previous case with $L_x\times L_y=101\times 11$ shown in Fig.~\ref{Fig.quench.from160}(c). However, as seen in Figs.~\ref{Fig.NLO2}(b1)-(b4), the order parameter has a two-dimensional pattern, which can be symbolically written as $\Delta_{\bm{r}_j}=\Delta[\cos(Q x_j)+\cos(Q y_j)]$, before the system relaxes to NLO. We will refer to the transient superconducting state characterized by the order parameter shown in Figs.~\ref{Fig.NLO2}(b1)-(b4) as two-dimensional NLO (2D-NLO).
\par
\begin{figure}[t]
\centering
\includegraphics[width=8.6cm]{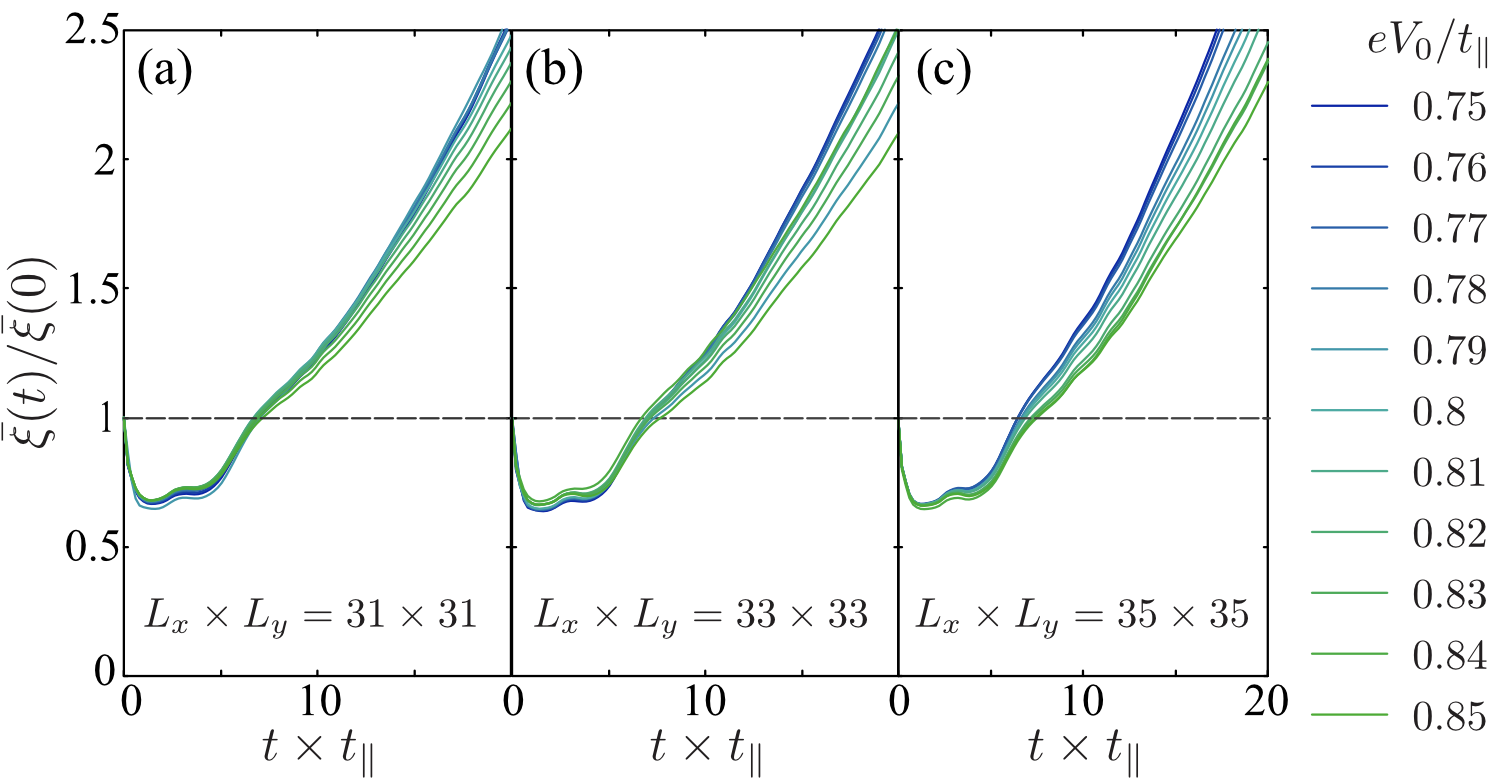}
\caption{Stability analysis of 2D-NLO for various voltages and system sizes. We set $\gamma/t_\parallel=0.01$ and $T_{\rm env}/t_\parallel=0.05$. We set (a) $L_x\times L_y = 31\times 31$, (b) $L_x\times L_y = 33 \times 33$, and (c) $L_x\times L_y = 35 \times 35$.}
\label{Fig.stability.detail}
\end{figure}
\par
Figure~\ref{Fig.NLO2}(b) indicates that the 2D-NLO is not a stable steady state. It should be noted, however, that we cannot rule out the possibility that the present instability of 2D-NLO is a finite-size effect. In calculations using a periodic lattice, the momentum $\bm{q}$ of the order parameter $\Delta_{\bm{q}}$ can only take values commensurate with the system size, which would be detrimental to 2D-NLO. To check this possibility, we perform the stability analysis of 2D-NLO for different voltages and system sizes. This is done by adding small superconducting fluctuations $\xi_j$ to the 2D-NLO order parameter at $t=0$ and investigating the time-evolution of the amplitude of the fluctuations, given by
\begin{equation}
\bar{\xi}(t) = \frac{1}{N}\sum_{j=1}^N|\xi_j(t)|.
\end{equation}
We set ${\rm Re} \xi_j(0)/\Delta_0$ and ${\rm Im} \xi_j(0)/\Delta_0$ as random numbers between $-0.1$ and $0.1$. If 2D-NLO is (un)stable,  $\bar{\xi}(t)$ decays (amplifies) as a function of time $t$.
\par
Figure~\ref{Fig.stability.detail} summarizes the results of the stability analysis. Since the frequency of the LO-type order parameter depends on the applied voltage, for some voltage and system size, the modulation of the 2D-NLO order parameter is expected to be almost commensurate with the system size. With this in mind, we judge from Fig.~\ref{Fig.stability.detail} that 2D-NLO is a metastable (linearly stable but non-linearly unstable) state. Thus, when the system enters region II in Fig.~\ref{Fig.phase.diagram}(a) from the normal phase, 2D-NLO initially appears, reflecting the four-fold rotational symmetry of the lattice potential, but the system would eventually relax to a more stable NLO, as shown in Fig.~\ref{Fig.NLO2}(b).
\par
We note that the possibility of the two-dimensional LO (2D-LO) has also been discussed in thermal equilibrium superconductivity under an external magnetic field \cite{Wang2006}. However, it is known that the unidirectional LO (just like NLO) is always energetically favored compared to 2D-LO in the two-dimensional attractive Hubbard model under an external magnetic field \cite{Wang2006}. 
\par
Before ending this section, we briefly comment on the overdamped behavior of the order-parameter amplitude $|\bar{\Delta}|$. When a system parameter is quenched, the order-parameter amplitude often oscillates while relaxing to a new steady-state value. A well-known example is the Higgs oscillation induced by the pairing interaction $U$ quench \cite{Collado2019}. However, when we quench the voltage $V$, $|\bar{\Delta}|$ relaxes to a new-steady value {\it without} oscillation, as shown in Figs.~\ref{Fig.quench.from000}, \ref{Fig.quench.from160}, and \ref{Fig.NLO2}(a). This is because unlike in the case of an interaction $U$ quench (where the system immediately feels the parameter change), the main system feels the voltage change only by exchanging electrons with the normal-metal leads \cite{Snyman2009}. Thus, for the voltage $V$ quench, $|\bar{\Delta}|$ cannot change to a new steady-state value faster than the electron exchange rate ($\sim 1/\gamma$) between the system and the normal-metal leads. Since this rate is comparable to the damping rate ($\sim 1/\gamma$), $|\bar{\Delta}(t)|$ shows an overdamped behavior for the voltage $V$ quench.
\par
\par
\section{Summary}
\par
To summarize, we have studied the nonequilibrium properties of a normal metal-superconductor-normal metal (N-S-N) junction consisting of a thin-film superconductor. When a bias voltage is applied between the normal-metal leads, the superconductor is driven out of equilibrium. By using the nonequilibrium Green's function technique, we derived a quantum kinetic equation for the nonequilibrium superconductor, to determine the superconducting order parameter self-consistently. The derived quantum kinetic equation is an integro-differential equation with memory effects. By utilizing a pole expansion of the Fermi-Dirac function, we converted the equation into ordinary differential equations, which are suitable for numerical calculations.
\par
By solving the quantum kinetic equation, we showed that the voltage-driven superconductor always relaxes to a certain nonequilibrium steady state. The resulting nonequilibrium phase diagram is presented in Fig.~\ref{Fig.phase.diagram}. In this phase diagram, a nonuniform superconducting state with a spatially oscillating order parameter $\Delta_{\bm{r}_j}=\Delta \cos(Q x_j)$, which is analogous to the Larkin-Ovchinnikov (LO) state in a superconductor under a magnetic field, is found in region II. We pointed out that the nonequilibrium LO (NLO) is induced by the nonequilibrium energy distribution of electrons in the superconductor, which has a two-step structure reflecting the different Fermi levels of the left and right normal-metal leads. We also found that the system exhibits bistability in regions II and III of the phase diagram. We showed that the bistability leads to hysteresis in the voltage-current characteristics of the junction.
\par
We end by listing some future problems. In the thermal-equilibrium case, the kernel polynomial method is known to be very useful in studying large-scale inhomogeneous superconductors \cite{Covaci2010, Nagai2012, Nagai2020}. The combination of the method and our kinetic equation may enable large-scale simulations to reexamine the stability of the two-dimensional NLO in region II. Applying our kinetic equation to a voltage-driven superconducting wire is an interesting future problem. In a voltage-driven superconducting wire, the time-periodic superconducting state associated with the appearance of the phase-slip centers is well known \cite{Langer1967, Skocpol1974, Kramer1977, Ivlev1984, Bezryadin2000, Vodolazov2003, Yerin2013, Yerin2013_2}, but our results suggest that not only temporally but also spatially inhomogeneous superconductivity would be realized due to the nonequilibrium energy distribution function having the two-step structure. The search for unconventional ordered phases in nonequilibrium quantum many-body systems is currently a rapidly evolving field, so that our results would contribute to the further development of this research field.
\par
\begin{acknowledgments}
\par
This work was supported by the Delta-ITP consortium and by the research program ``Materials for the Quantum Age" (QuMat). These are programs of the Netherlands Organisation for Scientific Research (NWO) and the Gravitation program, respectively, which are funded by the Dutch Ministry of Education, Culture and Science (OCW). T.K. was supported by MEXT and JSPS KAKENHI Grant-in-Aid for JSPS fellows Grant No. JP21J22452. Y.O. was supported by a Grant-in-aid for Scientific Research from MEXT and JSPS in Japan (No. JP19K03689 and No. JP22K03486).
\end{acknowledgments}
\par
\appendix
\begin{widetext}
\section{Derivation of the self-energy corrections \label{sec.App.self}}
\par
To derive the self-energy corrections $\hat{\bm{\Sigma}}_{\rm int}(t, t')$ and $\hat{\bm{\Sigma}}_{\rm lead}(t,t')$ given in Eqs.~\eqref{eq.selfR.int}, \eqref{eq.selfL.int}, \eqref{eq.selfR.env} and \eqref{eq.selfR.env}, we conveniently introduce the $4\times 4$ Nambu-Keldysh Green's function \cite{Yamaguchi2012, Hanai2017, Kawamura2022}, given by
\begin{equation}
\check{\bm{G}}_{jk}(t, t') =
\begin{pmatrix}
\bm{G}^{\rm r}_{jk}(t, t') & \bm{G}^{\rm k}_{jk}(t, t') \\[4pt]
0 & \bm{G}^{\rm a}_{jk}(t, t') 
\end{pmatrix}.
\label{eq.app.G.Keldysh}
\end{equation}
Here, $\bm{G}^{\rm r}_{jk}$ and $\bm{G}^{\rm a}_{jk}$ are given in Eq.~\eqref{eq.Gra} and 
\begin{equation}
\bm{G}^{\rm k}_{jk}(t, t') = -i
\begin{pmatrix}
[c_{j,\up}(t), c^\dagger_{k,\up}(t')]_- & 
[c_{j,\down}(t), c_{k,\up}(t')]_- \\
[c^\dagger_{j, \up}(t), c^\dagger_{k, \down}(t') ]_- & 
[c^\dagger_{j, \down}(t), c_{k, \down}(t')]_-
\end{pmatrix}
\end{equation}
is the Keldysh component. The lesser component in Eq.~\eqref{eq.G<} is related to $\bm{G}^{{\rm r}, {\rm a}, {\rm k}}_{jk}(t, t')$ as
\begin{equation}
\bm{G}^<_{jk}(t,t')=\frac{1}{2}\big[ \bm{G}^{\rm k}_{jk} -\bm{G}^{\rm r}_{jk} +\bm{G}^{\rm a}_{jk}\big](t,t').
\label{eq.rak<}
\end{equation}
\par
\subsection{Interaction effects $\hat{\bm{\Sigma}}_{\rm int}^{{\rm r}, {\rm a}, <}(t,t')$}
\par
\begin{figure}[t]
\centering
\includegraphics[width=8.2cm]{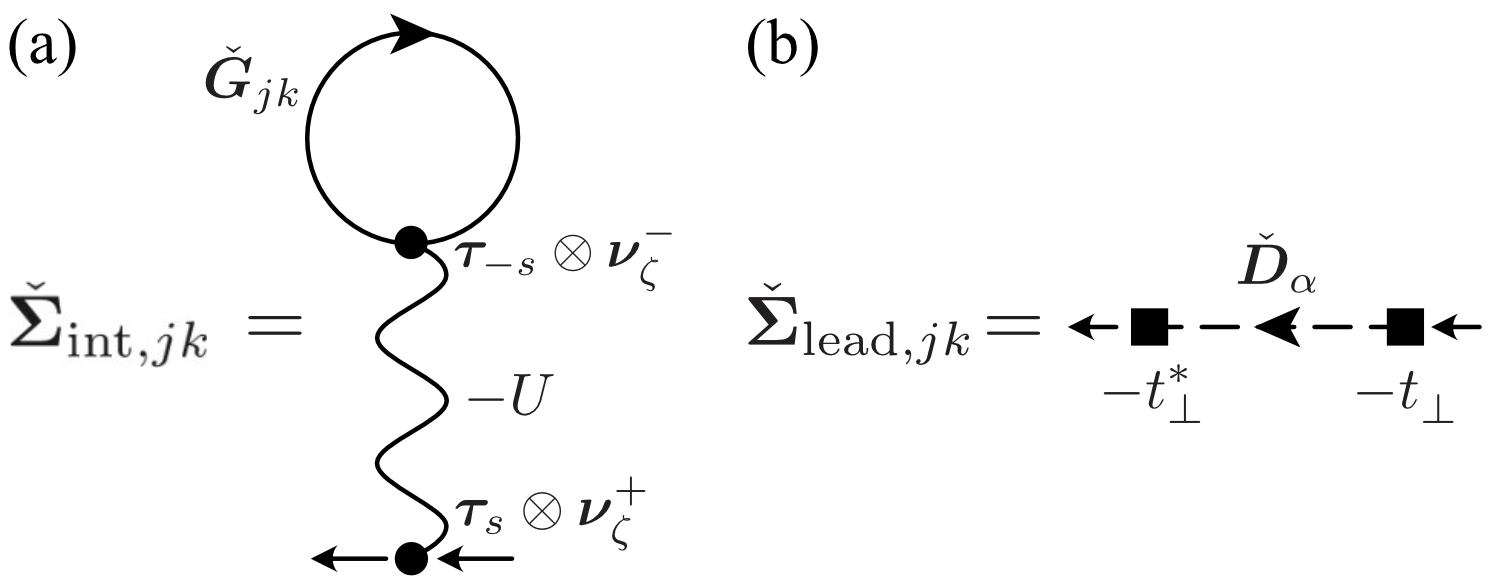}
\caption{Diagrammatic representation of the $4\times 4$ Nambu-Keldysh self-energy corrections. (a) $\check{\bm{\Sigma}}_{{\rm int}, jk}$ describes effects of the on-site pairing interaction $-U$ in the mean-field BCS approximation. The solid line is the dressed Nambu-Keldysh Green’s function $\check{\bm{G}}_{jk}$ given in Eq.~\eqref{eq.app.G.Keldysh}. The wavy line is the pairing interaction $-U$, being accompanied by the vertices $\bm{\tau}_s \otimes \bm{\nu}_\zeta^\pm$ at both ends. (b) $\check{\bm{\Sigma}}_{{\rm lead}, jk}$ describes the effects of the system-lead couplings in the second-order Born approximation with respect to the hopping amplitude $t_\perp$. The dashed line denotes the non-interacting Green’s function $\check{\bm{D}}_{\alpha={\rm L}, {\rm R}}$ in the $\alpha$ reservoir given in Eq.~\eqref{eq.app.Res.Keldysh}. The solid square represents the hopping amplitude $-t_\perp$ between the system and the $\alpha$ reservoir.
} 
\label{Fig.Diagram}
\end{figure}
\par
In the mean-field BCS approximation, the $4\times 4$ matrix self-energy $\check{\bm{\Sigma}}_{{\rm int}, jk}(t, t')$ due to the interaction $-U$ is diagrammatically drawn as Fig.~\ref{Fig.Diagram}(a), which gives \cite{Yamaguchi2012, Hanai2017, Kawamura2022}
\begin{align}
\check{\bm{\Sigma}}_{{\rm int}, jk}(t, t') 
&=
\begin{pmatrix}
\bm{\Sigma}^{\rm r}_{{\rm int}, jk}(t, t') & 
\bm{\Sigma}^{\rm k}_{{\rm int}, jk}(t, t') \\[3pt]
0 & \bm{\Sigma}^{\rm a}_{{\rm int}, jk}(t, t')
\end{pmatrix}
\notag\\[4pt]
&=
iU \sum_{s=\pm} \sum_{\zeta=1,2}
\big(\bm{\tau}_s \otimes \bm{\nu}^+_\zeta \big){\rm Tr}_{\rm N}{\rm Tr}_{\rm K} \Big[\big(\bm{\tau}_{-s} \otimes \bm{\nu}^-_\zeta \big) \check{\bm{G}}(t, t')\Big]
\delta(t-t') \delta_{j, k}
\notag\\[4pt]
&=
\frac{iU}{2} \sum_{s=\pm}
\begin{pmatrix}
\bm{\tau}_s {\rm Tr}_{\rm N}\big[\bm{\tau}_{-s}  \bm{G}^{\rm k}_{jk}(t,t')\big] & 
\bm{\tau}_s {\rm Tr}_{\rm N}\big[\bm{\tau}_{-s} \bm{G}^{\rm r}_{jk}(t,t') +\bm{\tau}_{-s} \bm{G}^{\rm a}_{jk}(t,t')\big] 
\\[3pt]
\bm{\tau}_s {\rm Tr}_{\rm N}\big[\bm{\tau}_{-s} \bm{G}^{\rm r}_{jk}(t,t') +\bm{\tau}_{-s} \bm{G}^{\rm a}_{jk}(t,t')\big] 
&
\bm{\tau}_s {\rm Tr}_{\rm N}\big[\bm{\tau}_{-s} \bm{G}^{\rm k}_{jk}(t,t')\big] 
\end{pmatrix}
\delta(t-t') \delta_{j, k},
\label{eq.app.self.int}
\end{align}
where $\bm{\tau}_\pm = [\bm{\tau}_1\pm i\bm{\tau}_2]/2$ and,
\begin{equation}
\bm{\nu}_\zeta^+=\frac{1}{\sqrt{2}} \bm{\sigma}_{2-\zeta}, \hspace{0.5cm}
\bm{\nu}_\zeta^-=\frac{1}{\sqrt{2}} \bm{\sigma}_{\zeta-1},
\end{equation}
are vertex matrices \cite{Kawamura2022} with $\bm{\sigma}_{j=1,2,3}$ being the Pauli matrices acting on the Keldysh space. ${\rm Tr}_{\rm N}$ and ${\rm Tr}_{\rm K}$ stand for taking the trace over the Nambu space and the Keldysh space, respectively. Noting the definition of the superconducting order parameter
\begin{equation}
\Delta_j(t)
= -iU G^<_{jj}(t, t)_{12},
= -\frac{iU}{2} G^{\rm k}_{jj}(t, t)_{12},
\end{equation}
we simplify Eq.~\eqref{eq.app.self.int} as
\begin{align}
\check{\bm{\Sigma}}_{{\rm int}, jk}(t, t')=
\begin{pmatrix}
-\bm{\Delta}_j(t) & 0 \\[3pt]
0 & -\bm{\Delta}_j(t)
\end{pmatrix}
\delta(t-t') \delta_{j,k}.
\label{eq.app.self.int2}
\end{align}
Here, $\bm{\Delta}_j(t)$ is given in Eq.~\eqref{eq.OP.Nambu}. We note that the off-diagonal components of $\check{\bm{\Sigma}}_{{\rm int}, jk}(t, t')$ identically vanish because $G^{\rm r(a)}_{jj}(t, t)_{12}=G^{\rm r(a)}_{jj}(t, t)_{21}=0$. From Eqs.~\eqref{eq.rak<} and \eqref{eq.app.self.int2}, we obtain the $2N\times 2N$ matrix self-energy corrections $\hat{\bm{\Sigma}}^{{\rm r}, {\rm a}, <}_{\rm int}(t, t')$ in Eqs.~\eqref{eq.selfR.int} and \eqref{eq.selfL.int}.
\par
\subsection{System-lead coupling effects $\hat{\bm{\Sigma}}^{{\rm r}, {\rm a}, <}_{\rm lead}(t, t')$}
\par
In the second-order Bron approximation with respect to the hopping amplitude $t_\perp$, the $4\times 4$ matrix self-energy $\check{\bm{\Sigma}}_{{\rm lead}, jk}(t, t')$ describing the couplings with normal-metal leads is diagrammatically drawn as Fig.~\ref{Fig.Diagram}(b). Evaluating this diagram, we obtain \cite{Yamaguchi2012, Hanai2017, Kawamura2022}
\begin{equation}
\check{\bm{\Sigma}}_{{\rm lead}, jk}(t, t') =
|t_\perp|^2 \sum_{\alpha={\rm L}, {\rm R}} \sum_{\bm{p}} \check{\bm{D}}_\alpha(\bm{p}, t, t')\delta_{j, k}. \label{eq.app.self.lead}
\end{equation}
Here, 
\begin{equation}
\check{\bm{D}}_{\alpha={\rm L}, {\rm R}} (\bm{p}, t, t')
=
\begin{pmatrix}
\bm{D}^{\rm r}_\alpha(\bm{p}, t, t') & \bm{D}^{\rm k}_\alpha(\bm{p}, t, t') \\[3pt]
0 & \bm{D}^{\rm a}_\alpha(\bm{p}, t, t')
\end{pmatrix}
\label{eq.app.Res.Keldysh}
\end{equation}
is the non-interacting Nambu-Keldysh Green's function in the $\alpha$ reservoir. From the Heisenberg equation of the field operator, one has \cite{Jauho1994}
\begin{align}
&
\bm{D}^{\rm r}_{{\rm L}({\rm R})}(\bm{p}, t, t')= 
\int_{-\infty}^\infty \frac{d\omega}{2\pi} 
\frac{1}{\omega + i\delta -[\ep_{\bm{p}}-\mu_{\rm env}]\bm{\tau}_3} 
e^{-i\omega(t-t')} \exp\left(\mp i\int_{t'}^t dt_1 \frac{eV(t_1)}{2} \bm{\tau}_3\right)
,\\[3pt]
&
\bm{D}^{\rm a}_{{\rm L}({\rm R})}(\bm{p}, t, t')= 
\int_{-\infty}^\infty \frac{d\omega}{2\pi} 
\frac{1}{\omega -i\delta -[\ep_{\bm{p}}-\mu_{\rm env}]\bm{\tau}_3} 
e^{-i\omega(t-t')} \exp\left(\mp i\int_{t'}^t dt_1 \frac{eV(t_1)}{2} \bm{\tau}_3\right)
\label{eq.app.self.lead.a} ,\\[3pt]
&
\bm{D}^{\rm k}_{{\rm L}({\rm R})}(\bm{p}, t, t')= -2\pi i \int_{-\infty}^\infty \frac{d\omega}{2\pi} \delta(\omega -\ep_{\bm{p}}+\mu_{\rm env}) \tanh\left(\frac{\omega}{2T_{\rm env}}\right) e^{- i\omega (t-t')}
\exp\left(\mp i\int_{t'}^t dt_1 \frac{eV(t_1)}{2} \bm{\tau}_3\right)
\label{eq.app.self.lead.K} ,\\[3pt]
&
\bm{D}^<_{{\rm L}({\rm R})}(\bm{p}, t, t')= 
2\pi i \int_{-\infty}^\infty \frac{d\omega}{2\pi} \delta(\omega -\ep_{\bm{p}}+\mu_{\rm env}) f(\omega) e^{-i\omega(t-t')}
\exp\left(\mp i\int_{t'}^t dt_1 \frac{eV(t_1)}{2} \bm{\tau}_3\right),
\label{eq.app.self.lead.<}
\end{align}
where $\omega_\pm=\omega \pm i\delta$, with $\delta$ being an infinitesimally small positive number. 
\par
Assuming the constant density of state $\rho_\alpha(\omega)\equiv \rho$ in the $\alpha$ reservoir and performing the $\bm{p}$ summation in Eq.~\eqref{eq.app.self.lead}, we have
\begin{align}
&
\bm{\Sigma}^{\rm r (a)}_{{\rm lead}, jk}(t, t') = \mp 2i\gamma \delta(t-t') \bm{\tau}_0 \delta_{j,k} ,\label{eq.app.self.lead.R}\\
&
\bm{\Sigma}^{\rm k}_{{\rm lead}, jk}(t, t') = 
-2i\gamma \sum_{\eta=\pm} 
\int_{-\infty}^\infty \frac{d\omega}{2\pi} e^{-i\omega (t-t')}
\tanh\left(\frac{\omega -\eta e V_0/2}{2T_{\rm env}}\right) \bm{\tau}_0 \delta_{j,k}\exp\left(-i \eta \int_{t'}^t dt_1 \frac{e\Delta V(t_1)}{2} \right).
\end{align}
Here, $\gamma$ is given in Eq.~\eqref{eq.muLR}. In deriving Eq.~\eqref{eq.app.self.lead.R}, we neglect the real part of the self-energy, which only gives the constant energy shift. The lesser component $\bm{\Sigma}^<_{{\rm lead}, jk}(t, t')$ is obtained from $\bm{\Sigma}^{{\rm r}, {\rm a}, {\rm k}}_{{\rm lead}, jk}(t,t')$ as 
\begin{align}
\bm{\Sigma}^<_{{\rm lead}, jk}(t, t') &=
\frac{1}{2}\big[\bm{\Sigma}^{\rm k}_{{\rm lead}, jk} -\bm{\Sigma}^{\rm r}_{{\rm lead}, jk} +\bm{\Sigma}^{\rm a}_{{\rm lead}, jk}\big](t, t')
\notag\\[4pt]
&=2i\gamma \sum_{\eta=\pm} 
\exp\left(-i \eta \int_{t'}^t dt_1 \frac{e\Delta V(t_1)}{2} \right)
\int_{-\infty}^\infty \frac{d\omega}{2\pi}e^{-i\omega(t-t')} f\left(\omega -\eta \frac{eV_{0}}{2}\right) \bm{\tau}_0 \delta_{j,k}. 
\label{eq.app.self.lead.L}
\end{align}
From Eqs.~\eqref{eq.app.self.lead.R} and \eqref{eq.app.self.lead.L}, we obtain the $2N\times 2N$ matrix self-energy corrections $\hat{\bm{\Sigma}}^{{\rm r}, {\rm a}, <}_{\rm lead}(t, t')$ in Eqs.~\eqref{eq.selfR.env} and \eqref{eq.selfL.env}.
\par
\section{Vanishing of $\hat{\bm{G}}^{<}_{\rm iso}(t, t')$ in Eq.~\eqref{eq.GL.ini**} \label{eq.sec.app.vanish.Gini}}
We first introduce the ``inverse" Green's functions $\overrightarrow{\hat{\bm{G}}}^{-1}_0(t)$ and $\overleftarrow{\hat{\bm{G}}}^{-1}_0(t)$ that obey,
\begin{align}
&
\overrightarrow{\hat{\bm{G}}}^{-1}_0(t)
\hat{\bm{G}}^{\rm r(a)}_0(t,t')=
\delta(t-t') \hat{\bm{1}},\\
&
\hat{\bm{G}}^{\rm r(a)}_0(t,t')
\overleftarrow{\hat{\bm{G}}}^{-1}_0(t') =
\delta(t-t') \hat{\bm{1}}.
\end{align}
From the Heisenberg equation of the field operator, these inverse Green’s functions are found to have the forms,
\begin{align}
&
\overrightarrow{\hat{\bm{G}}}^{-1}_0(t) =  i\overrightarrow{\partial}_t \hat{\bm{1}} -\hat{\bm{\m{H}}}_0,
\label{eq.app.G0inv.R}
\\
&
\overleftarrow{\hat{\bm{G}}}^{-1}_0(t') = -i\overleftarrow{\partial}_{t'} \hat{\bm{1}}
-\hat{\bm{\m{H}}}_0,
\label{eq.app.G0inv.L}
\end{align}
where $\hat{\bm{\m{H}}}_0$ is given in Eq.~\eqref{eq.H0} and the left (right) arrow on each differential operator means that it acts on the left (right) side of this operator. 
\par
The Dyson equation~\eqref{eq.Dyson.GR} (which is called the right Dyson equation) is known to be equivalent to the left Dyson equation \cite{RammerBook}, given by
\begin{equation}
\hat{\bm{G}}^{\rm r(a)}(t,t')=
\hat{\bm{G}}^{\rm r(a)}_0(t,t') +
\int_{-\infty}^\infty dt_1 \int_{-\infty}^\infty dt_2
\hat{\bm{G}}^{\rm r(a)}(t, t_1) \hat{\bm{\Sigma}}^{\rm r(a)}(t_1, t_2) \hat{\bm{G}}^{\rm r(a)}_0(t_2, t').
\end{equation}
Operating $\overrightarrow{\hat{\bm{G}}}^{-1}_0(t)$ and $\overleftarrow{\hat{\bm{G}}}^{-1}_0(t')$, respectively, to the right and left Dyson equation, we have
\begin{align}
&
\overrightarrow{\hat{\bm{G}}}^{-1}_0(t)\hat{\bm{G}}^{\rm a}(t,t')
=
\delta(t-t') \hat{\bm{1}} + 
\int_{-\infty}^\infty dt_1 \hat{\bm{\Sigma}}^{\rm a}(t, t_1)\hat{\bm{G}}^{\rm a}(t_1, t'),
\label{eq.app.B6*}
\\
&
\hat{\bm{G}}^{\rm r}(t,t')\overleftarrow{\hat{\bm{G}}}^{-1}_0(t') =
\delta(t-t') \hat{\bm{1}} + 
\int_{-\infty}^\infty dt_1 \hat{\bm{G}}^{\rm r}(t, t_1) \hat{\bm{\Sigma}}^{\rm r}(t_1, t').
\label{eq.app.B7*}
\end{align}
Using Eqs.~\eqref{eq.app.B6*} and \eqref{eq.app.B7*}, we can rewrite $\hat{\bm{G}}^{<}_{\rm iso}(t, t')$ in Eq.~\eqref{eq.GL.ini**} as
\begin{align}
\hat{\bm{G}}^{<}_{\rm iso}(t,t')
&=
\int_{-\infty}^\infty dt_1 \int_{-\infty}^\infty dt_2
\hat{\bm{G}}^{\rm r}(t,t_1)\overleftarrow{\hat{\bm{G}}}^{-1}_0(t_1) 
\hat{\bm{G}}_0^<(t_1,t_2) 
\overrightarrow{\hat{\bm{G}}}^{-1}_0(t_2)\hat{\bm{G}}^{\rm a}(t_2,t')
\notag\\[4pt]
&=
-i \left.\int_{-\infty}^\infty dt_2
\hat{\bm{G}}^{\rm r}(t,t_1) \hat{\bm{G}}_0^<(t_1,t_2) 
\overrightarrow{\hat{\bm{G}}}^{-1}_0(t_2)\hat{\bm{G}}^{\rm a}(t_2,t')
\right|_{t_1=-\infty}^{t_1=\infty}
\notag\\
&\hspace{4cm}+
\int_{-\infty}^\infty dt_1 \int_{-\infty}^\infty dt_2
\hat{\bm{G}}^{\rm r}(t,t_1)\overrightarrow{\hat{\bm{G}}}^{-1}_0(t_1) 
\hat{\bm{G}}_0^<(t_1,t_2) 
\overrightarrow{\hat{\bm{G}}}^{-1}_0(t_2)\hat{\bm{G}}^{\rm a}(t_2,t').
\label{eq.B7}
\end{align}
To obtain the second line, we have integrated by parts with respect to $t_1$. Since the Dyson equation~\eqref{eq.Dyson.GR} can be formally solved as 
\begin{equation}
\bm{G}^{\rm r}(t, t') = -i\Theta(t-t') e^{-2\gamma(t-t')}
\exp\left(-\int_{t}^{t'} dt_1 \hat{\bm{\m{H}}}_{\rm BdG}(t_1)\right),
\label{eq.app.Gr}
\end{equation}
the first term in Eq.~\eqref{eq.B7} vanishes. [Note that $\bm{G}^{\rm r}(t, +\infty)=0$ due to the step function $\Theta(t-t')$, and $\bm{G}^{\rm r}(t, -\infty)=0$ due to the damping factor $e^{-2\gamma(t-t')}$ arising from the system-lead couplings.] In Eq.~\eqref{eq.app.Gr}, $\hat{\bm{\m{H}}}_{\rm BdG}(t)$ is given in Eq.~\eqref{eq.HBdG}.
\par
For the second term in Eq.~\eqref{eq.B7}, from the Heisenberg equation of the field operator,
$\hat{\bm{G}}_0^<(t, t')$ is found to have the from,
\begin{equation}
\hat{\bm{G}}_0^<(t, t') = i e^{-i \hat{\bm{\m{H}}}_0(t-t')} \braket{\bm{\Psi}^\dagger \bm{\Psi}}_{H_0}.
\label{eq.app.G0L*}
\end{equation}
Here, $\hat{\bm{\Psi}}$ is the $2N$-component Nambu field given in Eq.~\eqref{eq.Nambu.field} and $\braket{\cdots}_{H_0}$ denotes the expectation value on the thermal equilibrium state at the temperature $T_{\rm iso}$ before the system-lead couplings, as well as the pairing interaction, are switched on. From Eqs.~\eqref{eq.app.G0inv.R} and \eqref{eq.app.G0L*},  one has
\begin{equation}
\overrightarrow{\hat{\bm{G}}}^{-1}_0(t)\hat{\bm{G}}_0^<(t, t') =0,
\end{equation}
and the second term in Eq.~\eqref{eq.B7} is also 
found to vanish. Thus, we find $\hat{\bm{G}}^{<}_{\rm iso}(t,t')=0$.
\section{Derivation of Eq.~\eqref{eq.main} \label{eq.sec.app.QKE}}
\par
Operating the inverse Green's function $\overrightarrow{\hat{\bm{G}}}^{-1}_0(t)$ in Eq.~\eqref{eq.app.G0inv.R} and $\overleftarrow{\hat{\bm{G}}}^{-1}_0(t')$ in Eq.~\eqref{eq.app.G0inv.L} to the Dyson equation \eqref{eq.Dyson.GL} from left and right, we have
\begin{align}
&
\overrightarrow{\hat{\bm{G}}}^{-1}_0(t)\hat{\bm{G}}^{<}(t,t')=
\big[
\hat{\bm{\Sigma}}^<  \circ \hat{\bm{G}}^{\rm a} + 
\hat{\bm{\Sigma}}^{\rm r} \circ \hat{\bm{G}}^<
\big](t, t')
\label{eq.left},\\
&
\hat{\bm{G}}^{<}(t,t')
\overleftarrow{\hat{\bm{G}}}^{-1}_0(t') =
\big[
\hat{\bm{G}}^{\rm r} \circ \hat{\bm{\Sigma}}^< + 
\hat{\bm{G}}^< \circ \hat{\bm{\Sigma}}^{\rm a}
\big](t, t').
\label{eq.right}
\end{align}
Here, we introduce the abbreviated notation, 
\begin{equation}
\big[\hat{\bm{A}}\circ \hat{\bm{B}} \big](t,t')=
\int_{-\infty}^\infty dt_1 \hat{\bm{A}}(t, t_1) \hat{\bm{B}}(t_1, t').
\end{equation}
In obtaining Eqs.~\eqref{eq.left} and \eqref{eq.right}, we used
\begin{align}
&
\overrightarrow{\hat{\bm{G}}}^{-1}_0(t)\hat{\bm{G}}^{\rm r}(t,t')
=
\delta(t-t') \hat{\bm{1}} + 
\big[\hat{\bm{\Sigma}}^{\rm r}\circ \hat{\bm{G}}^{\rm r}\big](t, t')
\label{eq.EOM.Gr}
,\\
&
\hat{\bm{G}}^{\rm a}(t,t')\overleftarrow{\hat{\bm{G}}}^{-1}_0(t') =
\delta(t-t') \hat{\bm{1}} + 
\big[\hat{\bm{G}}^{\rm a} \circ \hat{\bm{\Sigma}}^{\rm a}\big](t, t').
\end{align}
Subtracting Eq.~\eqref{eq.right} from Eq.~\eqref{eq.left} and setting $t=t'$, we obtain the equation of motion of the equal-time lesser Green's function $\hat{\bm{G}}^{<}(t)=\hat{\bm{G}}^{<}(t, t)$ as
\begin{align}
\overrightarrow{\hat{\bm{G}}}^{-1}_0(t)\hat{\bm{G}}^{<}(t)-
\hat{\bm{G}}^{<}(t)\overleftarrow{\hat{\bm{G}}}^{-1}_0(t)
&=
\big[
\hat{\bm{\Sigma}}^{\rm r} \circ \hat{\bm{G}}^< -
\hat{\bm{G}}^< \circ \hat{\bm{\Sigma}}^{\rm a} +
\hat{\bm{\Sigma}}^<  \circ \hat{\bm{G}}^{\rm a} -
 \hat{\bm{G}}^{\rm r} \circ \hat{\bm{\Sigma}}^< 
\big](t, t)
\equiv \hat{\bm{\m{I}}}(t),
\label{eq.KB}
\end{align}
where the collision term $\hat{\bm{\m{I}}}(t)=\hat{\bm{\m{I}}}_{\rm int}(t)+\hat{\bm{\m{I}}}_{\rm lead}(t)$ consists of the interaction term
\begin{equation}
\hat{\bm{\m{I}}}_{\rm int}(t) =
\big[
\hat{\bm{\Sigma}}^{\rm r}_{\rm int}\circ \hat{\bm{G}}^< -
\hat{\bm{G}}^< \circ \hat{\bm{\Sigma}}^{\rm a}_{\rm int} +
\hat{\bm{\Sigma}}^<_{\rm int}  \circ \hat{\bm{G}}^{\rm a} -
\hat{\bm{G}}^{\rm r} \circ \hat{\bm{\Sigma}}^<_{\rm int} 
\big](t, t'), \label{eq.I.int}
\end{equation}
as well as the system-lead coupling term
\begin{equation}
\hat{\bm{\m{I}}}_{\rm lead}(t) =
\big[
\hat{\bm{\Sigma}}^{\rm r}_{\rm lead} \circ \hat{\bm{G}}^< -
\hat{\bm{G}}^< \circ \hat{\bm{\Sigma}}^{\rm a}_{\rm lead} +
\hat{\bm{\Sigma}}^<_{\rm lead}  \circ \hat{\bm{G}}^{\rm a} -
\hat{\bm{G}}^{\rm r} \circ \hat{\bm{\Sigma}}^<_{\rm lead} 
\big](t, t'). \label{eq.I.lead}
\end{equation}
Substituting the self-energy corrections in Eqs.~\eqref{eq.selfR.int}, \eqref{eq.selfL.int}, \eqref{eq.selfR.env}, and \eqref{eq.selfL.env} to Eqs.~\eqref{eq.I.int} and \eqref{eq.I.lead}, one can evaluate each term as
\begin{align}
& \hat{\bm{\m{I}}}_{\rm int}(t) = -\big[ \hat{\bm{\Delta}}(t),  \hat{\bm{G}}^<(t) \big]_-
,\\[4pt]
& \hat{\bm{\m{I}}}_{\rm lead}(t) =
-4i\gamma \hat{\bm{G}}^<(t) -\hat{\bm{\Pi}}(t) -\hat{\bm{\Pi}}^\dagger(t),
\end{align}
where $\hat{\bm{\Pi}}(t)$ is given in Eq.~\eqref{eq.Pi}. Since the left-hand side of Eq.~\eqref{eq.KB} is evaluated as
\begin{equation}
\overrightarrow{\hat{\bm{G}}}_0(t)\hat{\bm{G}}^{<}(t)-
\hat{\bm{G}}^{<}(t)\overleftarrow{\hat{\bm{G}}}_0(t) =
i\partial_t \hat{\bm{G}}^{<}(t) - \big[\hat{\bm{\m{H}}}_0, \hat{\bm{G}}^{<}(t)\big]_-,
\end{equation}
we obtain Eq.~\eqref{eq.main}.
\par
\section{Pad\'{e} expansion of the Fermi-Dirac function \label{app.pole}}
In the Pad\'{e} expansion \cite{Ozeki2007, Karrasch2010}, the poles $\chi_{\eta, n}$ and residues $r_n$ in Eq.~\eqref{eq.Fermi.ex}
 are efficiently calculated by solving an eigenvalue problem $\bm{B} \ket{b_n} = b_n \ket{b_n}$ of a $2N_{\rm F} \times 2N_{\rm F}$ tridiagonal matrix $\bm{B}$ that has non-zero components
\begin{equation}
B_{j, j+1}= B_{j+1, j}=
\frac{1}{2\sqrt{[2j-1][2j+1]}}
\end{equation}
for $1\leq j \leq 2N_{\rm F}-1$ \cite{Karrasch2010}. Due to the symmetry of the matrix $\bm{B}$, all eigenvalues of $\bm{B}$ come in pairs $(b_{n}, -b_{n})$, where $b_{n=1,\cdots, N_{\rm F}}>0$. By using $b_n$ and $\ket{b_n}$, $\chi_n$ and $r_n$ are obtained as 
\begin{align}
& \chi_n = \frac{T_{\rm env}}{b_n}
,\\
& r_n =  \frac{|\braket{1|b_n}|^2}{4b_n^2},
\end{align}
where $\bra{1}=(1,0,\cdots,0)$. 
\par
The required number $N_{\rm F}$ of poles in Eq.~\eqref{eq.Fermi.ex} can be estimated from the superconducting order parameter $\Delta_j(0)$. As explained in Sec.~\ref{sec.NESS}, $\Delta_j(0)$ is self-consistently determined by solving the gap equation~\eqref{eq.gap.NESS} with $\hat{\bm{\Lambda}}$ in Eq.~\eqref{eq.def.Lamb}. Although the elements $\Lambda_{j=1,\cdots,2N}$ of $\hat{\bm{\Lambda}}$ are computed exactly as in Eq.~\eqref{eq.full.Lamb}, they can also be evaluated from the expansion of the Fermi-Dirac function in Eq.~\eqref{eq.Fermi.ex}. Substituting Eq.~\eqref{eq.Fermi.ex} into the first line in Eq.~\eqref{eq.full.Lamb} and using the residue theorem, one has 
\begin{align}
&
\Lambda^{N_{\rm F}}_j=\frac{i}{2}-\frac{iT_{\rm env}}{2} 
\sum_{\eta=\pm}\sum_{n=1}^{N_{\rm F}}\left[
\frac{r_n}{E_j -\chi_{\eta, n} -2i\gamma} +
\frac{r_n}{E_j -\chi^*_{\eta, n} +2i\gamma} \right].
\label{eq.app.Lamb}
\end{align}
Here, $\Lambda_j^{N_{\rm F}}$ in Eq.~\eqref{eq.app.Lamb} is reduced to $\Lambda_j$ in Eq.~\eqref{eq.full.Lamb} in the limit of $N_{\rm F}\to \infty$. Comparing $\Delta_j(0)$ obtained with $\Lambda_{j}^{N_{\rm F}}$ and $\Lambda_j$, we can estimate the required number $N_{\rm F}$ of poles to obtain the order parameter $\Delta_j$ with sufficient accuracy.
\par
\begin{figure}[t]
\centering
\includegraphics[width=8.2cm]{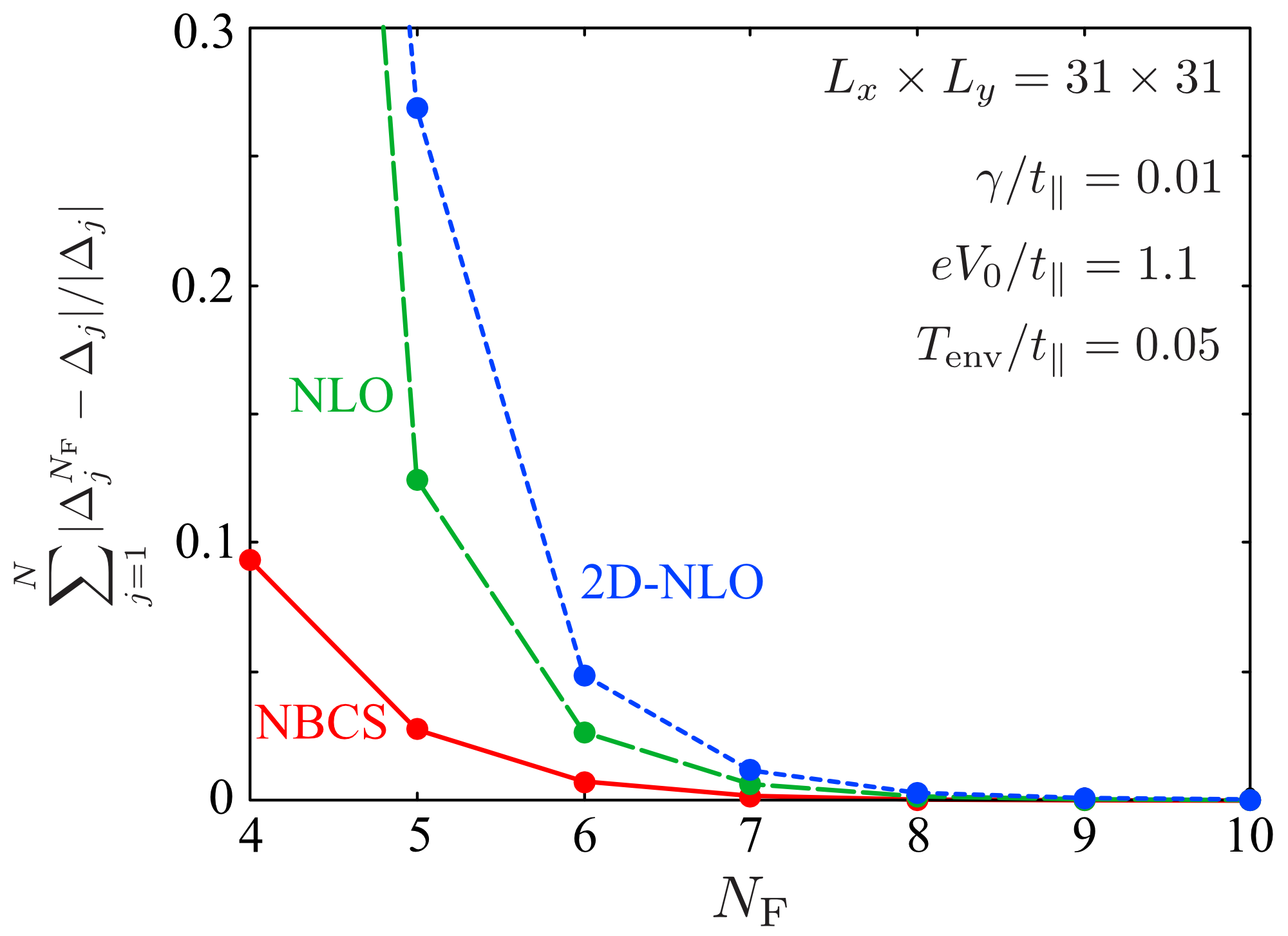}
\caption{Typical $N_{\rm F}$ dependence of $\sum_{j=1}^N|\Delta_j^{N_{\rm F}}-\Delta_j|/|\Delta_j|$ for several steady-state solutions. Here, $\Delta_j$ and $\Delta_j^{N_{\rm F}}$ are, respectively, obtained with Eq.~\eqref{eq.full.Lamb} and Eq.~\eqref{eq.app.Lamb}. We set $L_x=L_y=31$, $\gamma/t_\parallel=0.01$, $eV_0/t_\parallel=1.1$, and $T_{\rm env}/t_\parallel=0.05$.
}
\label{Fig.Err}
\end{figure}
\par
Figure~\ref{Fig.Err} shows the typical $N_{\rm F}$ dependence of
\begin{equation}
\sum_{j=1}^N \frac{|\Delta_j^{N_{\rm F}} -\Delta_j|}{|\Delta_j|}
\label{eq.err}
\end{equation}
for several steady-state solutions (NBCS, NLO, and 2D-NLO). For the details of these steady states, see Sec.~\ref{sec.result}.
In Eq.~\eqref{eq.err}, $\Delta_j$ and $\Delta_j^{N_{\rm F}}$ are, respectively, obtained with $\Lambda_j$ in Eq.~\eqref{eq.full.Lamb} and $\Lambda_j^{N_{\rm F}}$ in Eq.~\eqref{eq.app.Lamb}. As seen from this figure, the difference between $\Delta_j$ and $\Delta_j^{N_{\rm F}}$ becomes smaller as the number $N_{\rm F}$ of poles increases. In particular,  the difference between the two is less than $0.005$ $\%$ for all steady-state solutions, when $N_{\rm F}=10$. Thus, we choose $N_{\rm F}=10$ in our calculations.
\par
\section{Derivation of Eq.~\eqref{eq.noneq.gap} \label{sec.app.gapeq}}
\par
In momentum space, the $2\times 2$ matrix Nambu Green's functions are defined as \cite{Kawamura2022}
\begin{align}
\bm{G}^{\rm r}(\bm{k}, t,t') 
&= 
\big[	\bm{G}^{\rm a}(\bm{k}, t',t)\big]^\dagger
=
-i\Theta(t-t')
\begin{pmatrix}
\braket{[c_{\bm{k},\up}(t),  c^\dagger_{\bm{k},\up}(t')]_+} &  
\braket{[c_{\bm{k},\up}(t),  c_{-\bm{k}, \down}(t')]_+} \\
\braket{[c^\dagger_{-\bm{k},\down}(t),  c^\dagger_{\bm{k},\up}(t')]_+} & 
\braket{[c^\dagger_{-\bm{k},\down}(t),  c_{-\bm{k},\down}(t')]_+} 
\end{pmatrix}
,\\[4pt]
\bm{G}^<(\bm{k}, t,t') 
&= 
i
\begin{pmatrix}
\braket{c^\dagger_{\bm{k}, \up}(t') c_{\bm{k}, \up}(t)} & 
\braket{c_{-\bm{k}, \down}(t') c_{\bm{k}, \up}(t)} \\
\braket{c^\dagger_{\bm{k}, \up}(t') c^\dagger_{-\bm{k}, \down}(t)} &
\braket{c_{-\bm{k}, \down}(t') c^\dagger_{-\bm{k}, \down}(t)}
\end{pmatrix}.
\end{align}
When the system is in a NESS, the Nambu Green's functions $\bm{G}^{{\rm x}={\rm r}, {\rm a}, <}(t, t')$ depend only on the relative time $t-t'$. In frequency space, $\bm{G}^{\rm x}_{\rm NESS}(\omega)$ then obey the Keldysh-Dyson equations, 
\begin{align}
&
\bm{G}^{\rm r}_{\rm NESS}(\bm{k}, \omega)=
\bm{G}^{\rm r}_0(\bm{k}, \omega)	+
\bm{G}^{\rm r}_0(\bm{k}, \omega) 
\bm{\Sigma}^{\rm r}(\bm{k}, \omega) 
\bm{G}^{\rm r}_{\rm NESS}(\bm{k}, \omega),
\label{eq.app.Dyson.R}
\\[4pt]
&
\bm{G}^<(\bm{k}, \omega)=
\bm{G}^{\rm r}_{\rm NESS}(\bm{k}, \omega)
\bm{\Sigma}^<(\bm{k}, \omega)
\bm{G}^{\rm a}_{\rm NESS}(\bm{k}, \omega),
\label{eq.app.Dyson.L}
\end{align}
where
\begin{equation}
\bm{\Sigma}^{{\rm x}={\rm r}, {\rm a}, <}(\bm{k}, \omega) =
\bm{\Sigma}^{\rm x}_{\rm int}(\bm{k}, \omega)  + \bm{\Sigma}^{\rm x}_{\rm lead}(\bm{k}, \omega).
\label{eq.app.self.Nambu}
\end{equation}
In Eq.~\eqref{eq.app.self.Nambu}, $\bm{\Sigma}^{\rm x}_{\rm int}(\bm{k}, \omega)$ describes the interaction effects. In the mean-field BCS approximation, it is given by \cite{Kawamura2022}
\begin{align}
&\bm{\Sigma}^{\rm r}_{\rm int}(\bm{k}, \omega) =
\bm{\Sigma}^{\rm a}_{\rm int}(\bm{k}, \omega) = -\Delta \bm{\tau}_1
\label{eq.app.Nambu.sigR.int}
,\\[4pt]
& \bm{\Sigma}^{<}_{\rm int}(\bm{k}, \omega)=0,
\label{eq.app.Nambu.sigL.int}
\end{align}
with the Pauli matrices $\bm{\tau}_{j}$ acting on the Nambu space. Here, we assume a uniform superconducting order parameter ($\Delta_j =\Delta$), which is related to the off-diagonal component of the Nambu lesser Green's function as
\begin{equation}
\Delta = -iU \sum_{\bm{k}} \int_{-\infty}^\infty \frac{d\omega}{2\pi} G^<_{\rm NESS}(\bm{k}, \omega)_{12}.
\label{eq.OP.k}
\end{equation}
The system-lead coupling effects are summarized in $\bm{\Sigma}^{\rm x}_{\rm lead}(\bm{k}, \omega)$, which are given by \cite{Kawamura2022},
\begin{align}
&\bm{\Sigma}^{\rm r}_{\rm lead}(\bm{k}, \omega) =
\big[\bm{\Sigma}^{\rm a}_{\rm lead}(\bm{k}, \omega)\big]^\dagger=	-2i\gamma \bm{\tau}_0
\label{eq.app.Nambu.sigR.lead}
,\\[4pt]
& \bm{\Sigma}^{<}_{\rm lead}(\bm{k}, \omega) =
2i\gamma \left[f\left(\omega -\frac{eV_0}{2}\right) +f\left(\omega +\frac{eV_0}{2}\right)\right] \bm{\tau}_0.
\label{eq.app.Nambu.sigL.lead}
\end{align}
Substituting the self-energy corrections in Eqs.~\eqref{eq.app.Nambu.sigR.int}, \eqref{eq.app.Nambu.sigL.int}, \eqref{eq.app.Nambu.sigR.lead}, and \eqref{eq.app.Nambu.sigL.lead} into the Dyson equations~\eqref{eq.app.Dyson.R} and \eqref{eq.app.Dyson.L}, we have the Nambu Green's functions as
\begin{align}
&
\bm{G}^{\rm r(a)}_{\rm NESS}(\bm{k}, \omega) = 
\sum_{\eta=\pm} \frac{1}{\omega \pm 2i\gamma -E_{\bm{k}}} \bm{\Xi}^\eta_{\bm{k}}
\label{eq.k.Nambu.GR}
,\\[4pt]
&
\bm{G}^<_{\rm NESS}(\bm{k}, \omega) = 
\sum_{\eta=\pm} \frac{2i\gamma \big[f(\omega-eV_0/2)+f(\omega+eV_0/2)\big]}{[\omega-\eta E_{\bm{k}}]^2 +4\gamma^2} \bm{\Xi}^\eta_{\bm{k}}.
\label{eq.k.Nambu.GL}
\end{align}
Here, $E_{\bm{k}}$ is the Bogoliubov excitation energy given in Eq.~\eqref{eq.Bogo.Ep} and
\begin{equation}
\bm{\Xi}^{\eta=\pm}_{\bm{k}} = \frac{1}{2}\left[\bm{\tau}_0 \pm \frac{\ep_{\bm{k}}}{E_{\bm{k}}} \bm{\tau}_3 \mp \frac{\Delta}{E_{\bm{k}}} \bm{\tau}_1\right].
\end{equation}
Substituting Eq.~\eqref{eq.k.Nambu.GL} to Eq.~\eqref{eq.OP.k}, one has the nonequilibrium BCS gap equation~\eqref{eq.noneq.gap}.
\par
\section{Derivation of Eq.~\eqref{eq.Thouless} \label{sec.app.KM}}
\par
When the system is in the normal state ($\Delta_j=0$), the nonequilibrium Green's functions are given by
\begin{align}
G^{\rm r}_{\rm N}(\bm{k}, t, t')
&=
\big[G^{\rm a}_{\rm N}(\bm{k}, t', t)\big]^*
=
-i\theta(t-t') \braket{[c_{\bm{k},\sigma}(t),  c^\dagger_{\bm{k},\sigma}(t')]_+} 
,\\[4pt]
G^<_{\rm N}(\bm{k}, t, t')&=
i\braket{c^\dagger_{\bm{k}, \sigma}(t') c_{\bm{k}, \sigma}(t)}.
\end{align}
In a NESS, these Green's functions can easily be obtained from Eqs.~\eqref{eq.k.Nambu.GR} and \eqref{eq.k.Nambu.GL} by setting $\Delta =0$ and extracting the ($1,1$) component, which yields \cite{Kawamura2020JLTP, Kawamura2020}
\begin{align}
&
G^{\rm r(a)}_{{\rm N}, {\rm NESS}}(\bm{k}, \omega)= \frac{1}{\omega \pm 2i\gamma -\ep_{\bm{k}}} \label{eq.GNra}
,\\[4pt]
&
G^<_{{\rm N}, {\rm NESS}}(\bm{k}, \omega)=	\frac{-4i\gamma\big[1-f(\omega-eV_0/2)-f(\omega+eV_0/2)\big]}{[\omega-\ep_{\bm{k}}]^2 +4\gamma^2}. \label{eq.GN<}
\end{align}
We note that the lesser Green's function is related to the nonequilibrium momentum distribution $n_{\bm{k}}^{\rm neq}$ as
\begin{equation}
n_{\bm{k}}^{\rm neq}= \int_{-\infty}^\infty \frac{d\omega}{2\pi} G^<_{{\rm N}, {\rm NESS}}(\bm{k}, \omega).
\label{eq.nk.noneq}
\end{equation}
\par
Within the mean-field (ladder) approximation, the retarded particle-particle scattering $T$-matrix $\chi^{\rm r}(\bm{q}, \nu)$ can be evaluated as \cite{Kawamura2020JLTP, Kawamura2020}
\begin{equation}
\chi^{\rm r}(\bm{q}, \nu) = \frac{-U}{1 +U\chi^{\rm r}_0(\bm{q}, \nu)}.
\end{equation}
Here, $\chi^{\rm r}_0(\bm{q}, \nu)$ is the lowest-order pairing correlation function, given by
\begin{align}
\chi^{\rm r}_0(\bm{q}, \nu)&=
i\sum_{\bm{k}} \int_{-\infty}^\infty \frac{d\omega}{2\pi}
\Big[
G_{{\rm N}, {\rm NESS}}^<(\bm{k}+\bm{q}/2, \omega+\nu) 
G_{{\rm N}, {\rm NESS}}^{\rm r}(-\bm{k}+\bm{q}/2, -\omega)
\notag\\
&\hspace{6cm}+
G_{{\rm N}, {\rm NESS}}^{\rm r}(\bm{k}+\bm{q}/2, \omega+\nu) 
G_{{\rm N}, {\rm NESS}}^<(-\bm{k}+\bm{q}/2, -\omega)
\notag\\
&\hspace{6cm}+
G_{{\rm N}, {\rm NESS}}^{\rm r}(\bm{k}+\bm{q}/2, \omega+\nu) 
G_{{\rm N}, {\rm NESS}}^{\rm r}(-\bm{k}+\bm{q}/2, -\omega)
\Big].
\label{eq.chi0}
\end{align}
Substituting Eqs.~\eqref{eq.GNra} and \eqref{eq.GN<} into \eqref{eq.chi0}, we have Eq.~\eqref{eq.Thouless}.
\end{widetext}
\par


\begin{thebibliography}{99}
\bibitem{note.T} Strictly speaking, the temperature cannot be defined in the main system when it is out of equilibrium. In this paper, the term ``temperature" means $T_{\rm env}$ in the thermal-equilibrium normal-metal leads.
\end{thebibliography}
\end{document}